\documentclass[aps,superscriptaddress,english,12pt,nofootinbib]{revtex4}
\usepackage{float}
\DeclareUnicodeCharacter{2212}{-}
\usepackage{amssymb, amsmath, bm, dcolumn, epsf, graphicx, latexsym, slashed, simplewick}
\usepackage[utf8,utf8x]{inputenc}
\usepackage[normalem]{ulem}
\usepackage{color}
\usepackage{float}
\usepackage{natbib}
\usepackage[toc,title,page]{appendix}

\def\be{\begin{equation}}
\def\ee{\end{equation}}
\def\bea{\begin{eqnarray}}
\def\eea{\end{eqnarray}}

\usepackage[bookmarks=false]{hyperref}
\usepackage[capitalize]{cleveref}
\usepackage{comment}
\usepackage{silence}
\usepackage[normalem]{ulem}

\newcommand{\blue}{\textcolor{blue}} 

\begin{document}

\widetext

\title{Theoretical Priors and the Dark Energy Equation of State}
\author{Ido Ben-Dayan}
\author{Utkarsh Kumar}

\affiliation{Physics Department, Ariel University, Ariel 40700, Israel}
\date{\today}

\begin{abstract}
We revisit the theoretical priors used for inferring Dark Energy (DE) parameters. Any DE model must have some form of a tracker mechanism such that it behaved as matter or radiation in the past. Otherwise, the model is fine-tuned. We construct a model-independent parametrization that takes this prior into account and allows for a relatively sudden transition between radiation/matter to DE behavior. We match the parametrization with current data, and deduce that the adiabatic and effective sound speeds of DE play an important role in inferring the cosmological parameters. We find that there is a preferred transition redshift of $1+z\simeq 29-30$, and some reduction in the Hubble and Large Scale Structure tensions. 
\end{abstract}

\maketitle

\section{Introduction}
The standard $\Lambda$ cold dark matter ($\Lambda$CDM), also known as the Concordance Model, is a well-established cosmological model supported by data from numerous observations. The $\Lambda$CDM parameters have been constrained using the cosmic microwave background (CMB) measurements \cite{Planck:2018lbu,Planck:2018vyg,Planck:2019nip}, supernovae type Ia \cite{Pan-STARRS1:2017jku}, weak lensing \cite{DES:2021wwk,Hildebrandt:2016iqg,Hildebrandt:2018yau,HSC:2018mrq}, and galaxy clustering measurements \cite{Ross:2014qpa,Beutler:2011hx,BOSS:2016wmc,Bautista:2020ahg,Gil-Marin:2020bct,eBOSS:2020yzd,Neveux:2020voa,Hou:2020rse,duMasdesBourboux:2020pck} with about a percent accuracy. Despite its remarkable success, the validity of the Concordance Model is under investigation, as accumulation of data results in tensions between various measurements.
The last parameter that was added to the Concordance Model, was the cosmological constant, $\Lambda$. The presence of a dominant positive cosmological constant explains the present acceleration of the Universe. Except its relative energy density, $\Omega_{\Lambda}$, it should manifest itself in observations by an equation of state (eos) $w=-1$. This minimal $\Lambda$CDM model is still an excellent fit to data with $\Omega_{\Lambda}\simeq0.70$. Considering a constant eos without imposing $w=-1$, the data further constrains the eos to be $-1.14<w<-0.94$ \cite{Planck:2018vyg,Escamilla:2023oce}\footnote{This model is usually dubbed $w$CDM.}. A true constant, based on zero modes quantum fluctuations of the fields present in Nature, is expected to be many orders of magnitude above the observed value and there have been many attempts to reconcile the measurement with the theoretical expectation \cite{Weinberg:1988cp,Ben-Dayan:2021ayq,Ben-Dayan:2015nva,Peebles:2002gy,Carroll:1991mt,Carroll:2000fy,Weinberg:2000yb,Sahni:2002kh,Padmanabhan:2002ji,Nobbenhuis:2004wn,Polchinski:2006gy}.  This unappealing mismatch between theory and observations prompted the idea that the present acceleration is due to an evolving (usually scalar) field, dubbed Dark Energy (DE) \cite{Copeland:2006wr,Oks:2021hef}. Generically, this scalar field is not solving the so-called "old" cosmological constant problem (of the expected zero modes contribution to the energy density of the Universe). 
Nevertheless, assuming this question is somehow settled, the DE gives predictions for the present acceleration of the Universe.
If DE is realized in Nature then generically the eos is time/redshift dependent $w(z)$. Furthermore, the DE fluid/field will have fluctuations that will affect the growth of structure \cite{Huterer:2013xky,Caldera-Cabral:2009hoy,Ferlito:2022mok,Nunes:2021ipq}, thus providing a testable framework. Focusing on the eos, there are many possible parametrizations \cite{Chevallier:2000qy,Linder:2002et,Jassal:2005qc,Barboza:2008rh,Gerke:2002sx,Escamilla:2023shf,Adil:2021zxp,Aviles:2012ay,Sharma:2022ifr}, perhaps most notably the CPL parametrization \cite{Chevallier:2000qy,Linder:2002et}
\begin{equation}
    w(z)\simeq w_0+w_1\frac{z}{1+z},
\end{equation}
which is not valid at all redshifts.
Combining Planck with BAO and Supernovae data then gives $w_0= -0.957 \pm 0.080,w_1= -0.29 ^{+0.32}_{-0.26}$ \cite{Planck:2018vyg}.
Since DE is evolving with time, another tuning problem arises - why should its energy density and eos be such that it behaves nearly as a cosmological constant today \cite{Velten:2014nra,Steinhardt:1999nw,Zlatev:1998tr}? To avoid this coincidence, DE models are generically endowed with a tracker mechanism. The DE tracks the radiation or matter throughout the evolution of the Universe, until it decouples and acts as DE today \cite{Barreira:2011qi,Amendola:1999er}. We would like to include this theoretical prior into the parameter estimation analysis.

 In addition to theoretical difficulties, the values of some cosmological parameters inferred from different cosmological and astrophysical data are in tension with the Planck 2018 parameters for $\Lambda$CDM \cite{Planck:2018vyg}. Perhaps the most intriguing tensions are the Hubble $H_0$ and Large Scale Structure $S_8$ tension. The Hubble tension arises from the discrepancy in the measurement of the present value of the Hubble parameter between the model-dependent and model-independent probes. Currently, there is a $\sim 5 \sigma$ discrepancy between the SH0ES (model-independent) \cite{Riess:2016jrr,Riess:2018kzi,Riess:2019cxk,Riess:2020fzl,Riess:2021jrx,Freedman:2017yms,Freedman:2019jwv,Freedman:2020dne,Freedman:2021ahq,Camarena:2023rsd,Camarena:2019rmj,Camarena:2018nbr,Camarena:2019moy,Aviles:2016wel} and Planck 2018 CMB (model-dependent) measurements \cite{Planck:2018vyg}. However, in addition to the aforementioned measurements, the discrepancy in present day expansion rate persists when considering other experiments and data sets, see \cite{Verde:2019ivm} and references therein.

$S_8 = \sigma_8\sqrt{\Omega_m/0.3}$ is a parameter measuring linear fluctuations, where $\sigma_8 $ is the amplitude of linear fluctuations smoothed over $8 \text{Mpc} \, h^{-1}$, and $\Omega_m$ is the relative matter density today. In inferring $S_8$ there again seems to be a $2-3\sigma $ discrepancy between Planck and Weak Lensing experiments, e.g. \cite{DiValentino:2020vvd,White:2021yvw,Heymans:2020gsg, DES:2021zxv, Garcia-Garcia:2021unp, LSSTDarkEnergyScience:2022amt, Aiola:2020azj,Chen:2022jzq,Planck:2018vyg,Brout:2022vxf}. Various solutions have been proposed to solve both tensions individually \cite{Mortsell:2018mfj,Kamionkowski:2022pkx,Agrawal:2019lmo,Lin:2019qug,Smith:2019ihp,Alexander:2022own,Knox:2019rjx,Sabla:2021nfy,Sakstein:2019fmf,CarrilloGonzalez:2020oac,Hill:2020osr,Ivanov:2020ril,DiValentino:2020vvd,DiValentino:2018gcu,Cuceu:2019for,Schoneberg:2019wmt,White:2021yvw,Krolewski:2021znk,Verde:2019ivm,Riess:2020fzl,Riess:2021jrx,Vagnozzi:2019ezj,Vagnozzi:2020dfn,Vagnozzi:2021gjh,Vagnozzi:2021tjv,Feng:2017nss,Benetti:2017gvm,Kumar:2017dnp,Vagnozzi:2017ovm,Zhao:2017cud,Prilepina:2016rlq,Chacko:2016kgg,Ko:2016fcd,Xia:2016vnp,Kumar:2016zpg,Karwal:2016vyq,Ko:2016uft,Tram:2016rcw,Huang:2016fxc,DiValentino:2016hlg,SolaPeracaula:2016qlq,Sola:2016jky,Sola:2015wwa,Berezhiani:2015yta,Wojtak:2022bct,Freedman:2021ahq,Mortsell:2021tcx,Mortsell:2021nzg,Efstathiou:2020wxn,Hu:2023jqc,DiValentino:2022fjm,Abdalla:2022yfr,Shah:2021onj,Perivolaropoulos:2021jda,DiValentino:2021izs,DiValentino:2020zio,Oks:2021hef,DiValentino:2019ffd,DiValentino:2019jae,Dhawan:2021mel,Philcox:2021kcw,Zhang:2021yna,Yuan:2022jqf, Zhai:2022yyk, Simon:2022lde,DES:2021bvc, Busch:2022pcx,Planck:2015lwi,Krolewski:2021yqy, DES:2021zxv, Garcia-Garcia:2021unp, LSSTDarkEnergyScience:2022amt, Aiola:2020azj,Brout:2022vxf,Heymans:2020gsg,Schoneberg:2021qvd, Amon:2022azi,Sarkar:2023vpn,Sarkar:2023cpd,Bargiacchi:2023rfd,Dainotti:2023bwq,Bargiacchi:2021hdp,Bargiacchi:2023jse,Lenart:2022nip,Dainotti:2022bzg,Dainotti:2021pqg,Bargiacchi:2023rfd,Dainotti:2023bwq,Bargiacchi:2021hdp,Dainotti:2023ebr,Bargiacchi:2023jse,Lenart:2022nip,Dainotti:2022bzg,Dainotti:2021pqg}. Some solutions resolving the Hubble tension seem to worsen the LSS tension or vice-versa \cite{Reeves:2022aoi,Nunes:2021ipq,Dhawan:2021mel,DiValentino:2019jae,DiValentino:2019ffd,Vagnozzi:2019ezj}. A theory that addresses both of these conflicts at once would undoubtedly be appealing, so in this study we make a point of attempting to handle the $S_8$ and $H_0$ tensions concurrently \cite{Ben-Dayan:2023rgt,Artymowski:2021fkw,Artymowski:2020zwy}.

Recently, we have suggested an emerging DE model. In this model the DE is not a fundamental scalar field, but rather a thermodynamical collective behavior \cite{Ben-Dayan:2023rgt,Artymowski:2020zwy,Artymowski:2021fkw}. This approach solves the fine tuning and initial conditions problems, is free of the swampland conjectures \cite{Agrawal:2019dlm,Vafa:2005ui,Palti:2019pca,Obied:2018sgi,Agrawal:2018own,Garg:2018reu,Ooguri:2018wrx,Ben-Dayan:2018mhe} and does not modify gravity. The model has a built in tracker mechanism since it asymptotes to $w=-1$ at future infinity and to $w=1/3$ in the past - i.e. the fluid behaves as radiation in the past and transitions to DE behavior at some redshift. Our recent analysis shows that the model is restoring cosmological concordance and alleviating both the Hubble tension and the $S_8$ tension, performing significantly better than $\Lambda$CDM \cite{Ben-Dayan:2023rgt}.
Motivated by this success, we want to investigate whether this behavior is more generic.  

Since any valid DE model has $w(z=0)\simeq -1$ and any model with a tracker mechanism $w(z\gg 1)=1/3 \, \textit{or} \,\,  0$, we can implement this understanding into our parametrization. We can then test the novel parametrization and its effect on existing tensions such as the Hubble or $S_8$ tensions. We develop a phenomenological approach where 
at the most economical level is:
\begin{eqnarray}
 w_{DE} (a) = -1 + \frac{w_a}{ 1 + \left( a / a_t\right)^{n}},
\end{eqnarray} 
where $n$ is some integer, $a_t$ is a transition scale factor around which $w_{DE}$ transitions from $w_{DE}\simeq -1$ to $w_{DE}\simeq -1+w_a$. To demonstrate our approach, we shall consider the possibility $w_a=4/3$, such that the DE behaves as radiation at early times. Thus, the parameter that has to be inferred from measurements is neither $w_0$ nor $w_a$, but rather the transition redshift/scale factor $a_t$.  In this minimal approach $c_s^2$ is determined. So we are fitting less parameters that the usual CPL parametrization. We then extend our model to include $c_s$ and later also $n$ as a free parameters. We calculate the background and perturbations evolution and match to existing data.
We then perform a likelihood analysis of our model combining various data sets. We find some modest reduction in the Hubble and $S_8$ tension and some reduction in $\Delta \chi^2\sim -2--3$ compared to $\Lambda$CDM. The more interesting result is that we find the transition redshift to be highly constrained, $z_t=29-30$, providing a definite interval for exploration.

The paper is organized as follows. We first describe the details of background and perturbation evaluation of our proposed phenomenological fluid parametrization in section \ref{sec:model}. Next, in section \ref{sec:data} we discuss different data sets and methods used to analyze our proposed parametrization. This section includes the nomenclature of different models used, combinations of datasets and measures the asses the $H_0$ and $S_8$ tensions. Sections \ref{sec:results} deals with the results and finally we conclude in section \ref{sec:conclusions}

\section{Phenomenological Fluid Dark Energy} \label{sec:model}
We propose a scenario consisting of the standard cosmological model with cold dark matter (CDM) and DE using a phenomenological fluid with several critical key features which differ from the cosmological constant (CC). In this section will discuss the model, its background, and perturbative dynamics. Using our approach, we also illustrate the implications on cosmological tensions, such as Hubble ($H_0$) and Large Scale Structure $S_8$.
\subsection{Theory} Our proposed scenario mimics $\Lambda$, which is used to account for Dark Energy in late times and transits to radiation at early times. The model is specified by a time-varying equation of state for such phenomenological fluid. 
\paragraph*{\textbf{Conditions of DE model}} We start our discussion by pointing out the conditions for a successful Dark Energy Model as 
\begin{itemize}
    \item The Dark Energy model should be able to explain cosmic coincidence and the hierarchy problem of DE, e.g., $\Lambda$.
    \item There should be an era of dust domination, which is essential for the structure of the Universe.
    \item The equation of state of DE component today, defined as $w_{DE} = \frac{p_{DE}}{\rho_{DE}}$ where $\rho_{DE}$ and $p_{DE}$ being the energy density and pressure of the DE fluid respectively, must lie within $-1.14 < w_{DE} < -0.94$ \cite{Planck:2018vyg,Escamilla:2023oce}. 
    \item The evolution of DE fluid and hence the whole background and perturbative evolution of the Universe should be free from instability. Thus, there should be no gradient instability and the sound speed of DE fluid should be subluminal, i.e., $0< c_{DE, a}^2< 1$. Here $c_{DE, a}^2 = \frac{\dot{p}_{DE}}{\dot{\rho}_{DE}}$ is the squared sound speed of DE fluid. 
\end{itemize}
\paragraph*{\textbf{Background}}
The time-varying eos of a phenomenological fluid ($w_{DE}$) is governed by the following form
\begin{eqnarray}
 w_{DE} (a) = -1 + \frac{w_a}{ 1 + \left( a / a_t\right)^{n}}\, , \label{eq:eosde}
\end{eqnarray} 
 where we normalize the scale factor $a$ to be unity today. Equation (\ref{eq:eosde}) asymptotes to $w_{DE} (a) = -1 + w_a$ and  $w_{DE} (a) = -1 $ at early and late times respectively. $w_a$ tunes the desired eos of state of the Dark Energy fluid at early times, and $a_t$ is the "transition scale factor" or redshift at which the fluid approximately crosses $w_{DE} = 0$. The parameter $n$ controls the sharpness of this transition.  Using the equation \eqref{eq:eosde}, it is straightforward to derive the energy density of the fluid as 
\begin{eqnarray}
    \rho_{DE}(a) = \rho_{DE,0}\,a^{-3\,w_a}\,\left(\frac{1 + \left( a / a_t\right)^{n}}{1 + \left( 1 / a_t\right)^{n}}\right)^{3\,w_a / n}. \label{eq:ede}
\end{eqnarray}
Equations (\ref{eq:eosde}) and (\ref{eq:ede}) give us a full picture of the background for our Dark Energy model as a phenomenological fluid with up to $3$ parameter extension to $\Lambda$CDM.

Our proposal for Dark Energy using \textbf{P}henomenological \textbf{F}luid \textbf{D}ark \textbf{E}nergy (PFDE) has a built-in tracker mechanism that tracks the background throughout evolution. Since we would like the PFDE to track the background evolution of the Universe as a whole, we require that the eos of phenomenological fluid will lie within $\in [0\, ,1/3]$ at early times, which constrains $w_a$ as 
\begin{eqnarray}
    w_{DE}^{Early} \approx -1 + w_a\, \in[0\,, 1/3]\Rightarrow w_a \in [1\,, 4/3] 
\end{eqnarray}

We think the new parametrization has several advantages over existing ones: First, it captures the essence of DE with $w\simeq -1$. Second, it captures the tracker behavior of $w=1/3$ at large redshift. Third, there is a single free parameter - the transition redshift or scale factor $a_t$. Thus, in terms of the number of cosmological parameters it is equivalent to wCDM with fixed $w$. Later, when we allow for an arbitrary $c_s^2$, we have the same number of parameters as the CPL one, but with different theoretical input. 

\begin{figure}[!ht]
    \centering
    \includegraphics[scale=0.63]{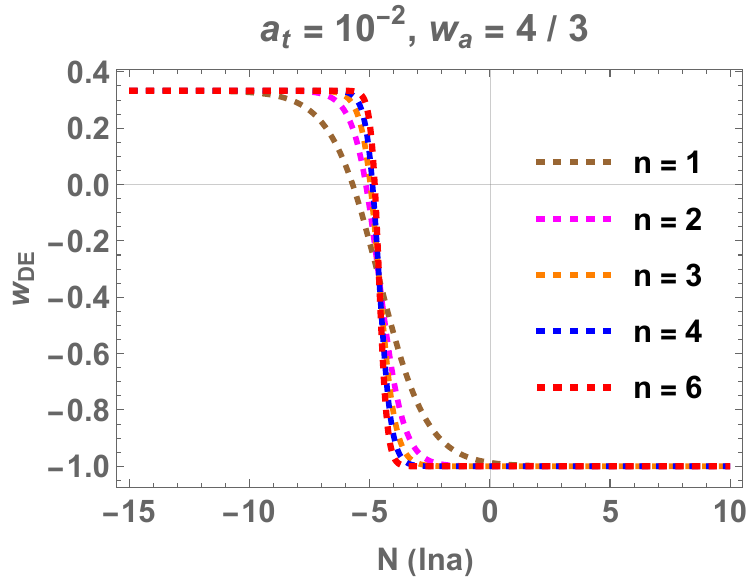}
    \includegraphics[scale=0.63]{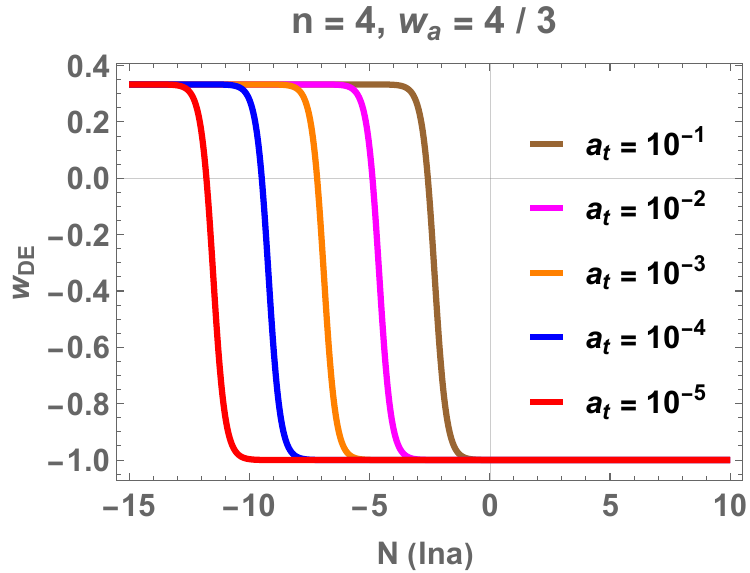}
    \caption{Evolution of Dark Energy equation of state $w_{DE}$ as a function of e-folds, $N=\ln a$, for different values of $n$ and $a_t$ while fixing other parameters in left and right panels respectively.}
    \label{fig:wde}
\end{figure}

Let us demonstrate the behavior of the model. First, we present examples of the effect of the steepness of transition and transition redshift on the DE eos $w_{DE}$ in Figure \ref{fig:wde}. It is clear from the plots that $w_{DE}$ transits from $1/3$ to -1, with different transition steepness (from varying $n$) and at different transition redshifts (varying $a_t$), at early and late times.

\begin{figure}[!ht]
    \centering
    \includegraphics[scale=0.63]{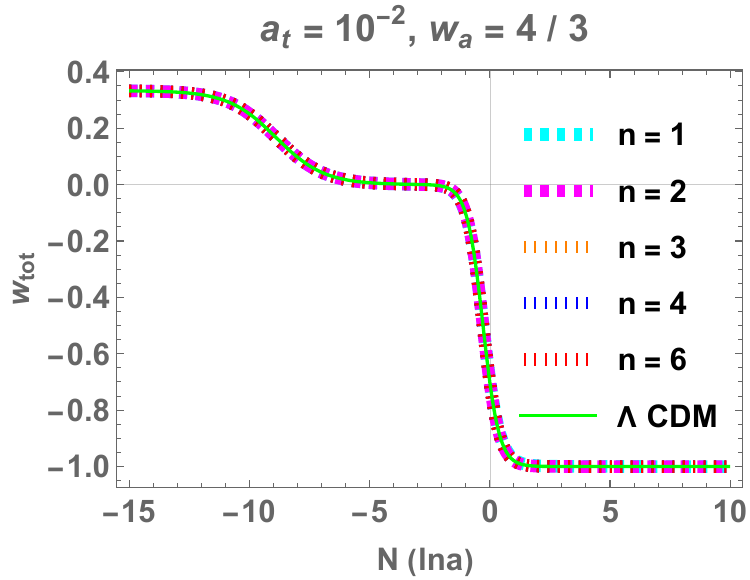}
    \includegraphics[scale=0.63]{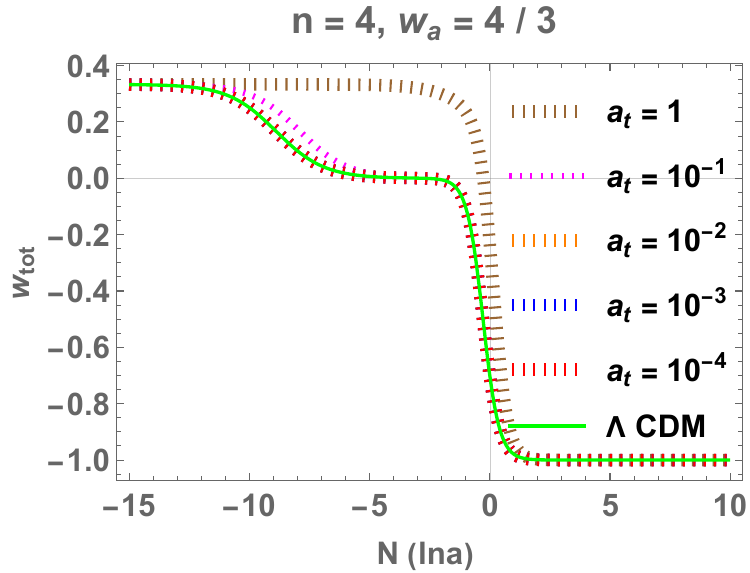}
    \caption{Evolution of total equation of state ($w_{tot}$) 
 as a function of e-folds, $N=\ln a$, for different values of $n$ and $a_t$ while fixing other parameters in left and right panels respectively.}
    \label{fig:wtot}
\end{figure}

\begin{figure}[!ht]
    \centering
    \includegraphics[scale=0.63]{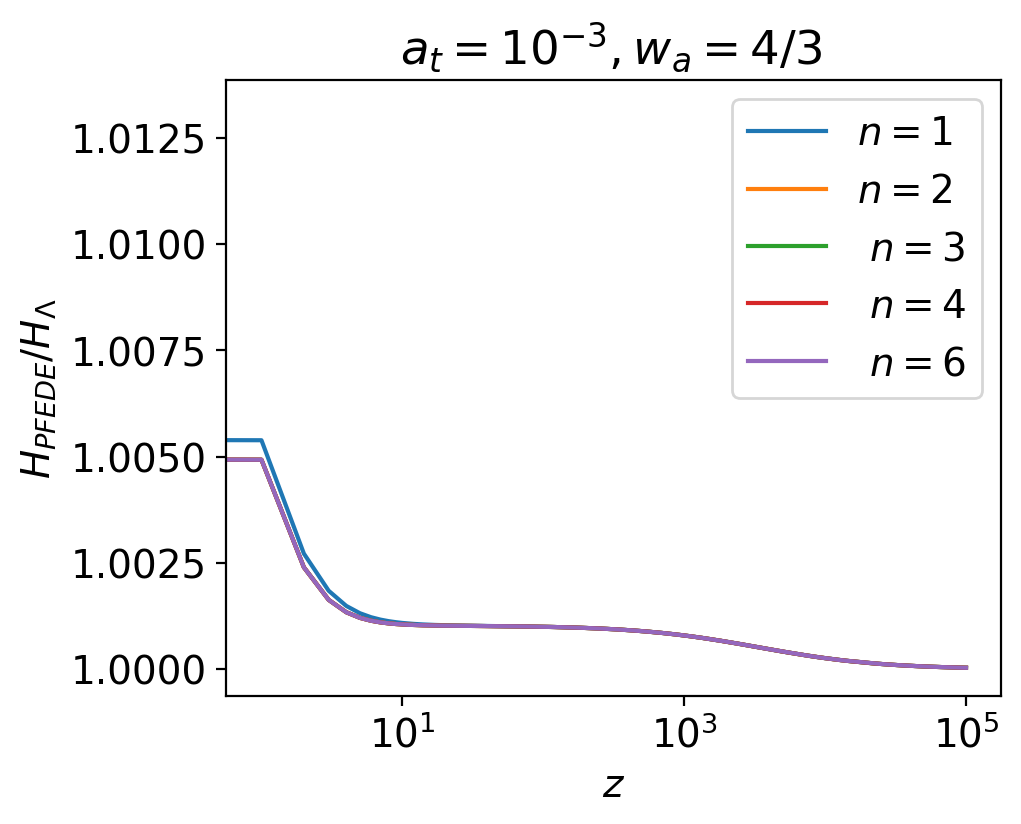}
    \includegraphics[scale=0.63]{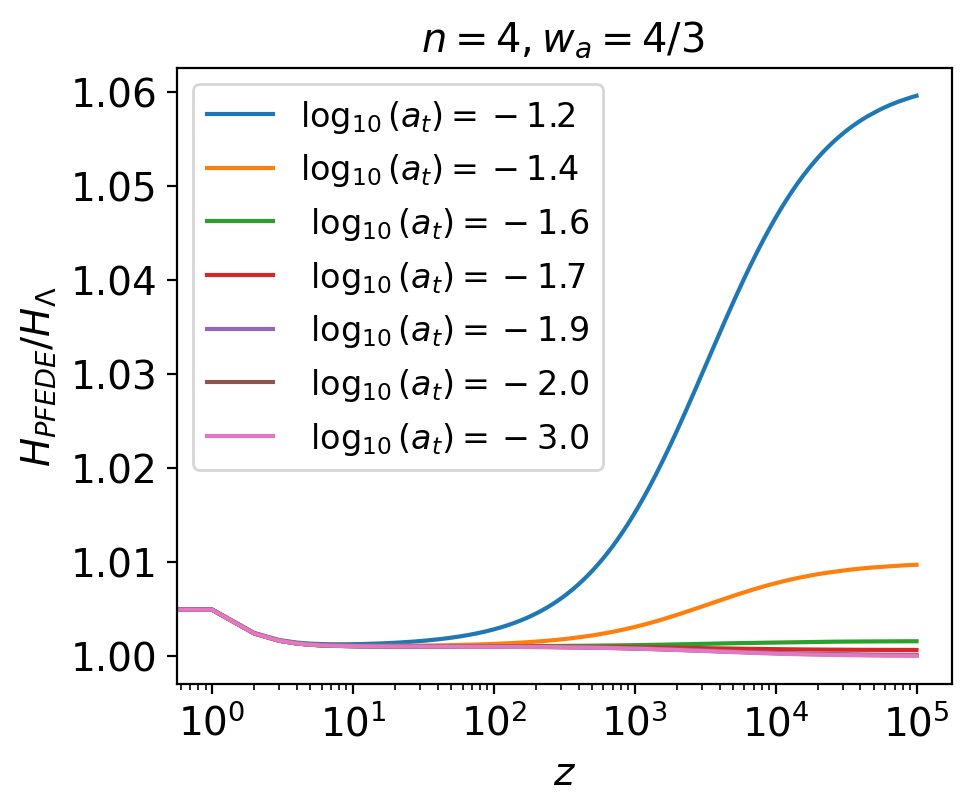}
    \caption{Comparison of the expansion rate of PFDE model and $\Lambda$CDM using their best fir parameters inferred from All data sets. We plot the $H_{PFDE} / H_{\Lambda}$ with varying $n$ and $a_t$ in the left and right panels respectively.}
    \label{fig:Hcom}
\end{figure}

Next, we discuss the evolution of the effective eos of the Universe, $w_{tot}$. There is almost no difference from the standard $\Lambda$CDM for different values $n$ while there is a strong dependence of $a_t$ on $w_{tot}$ as shown in the left and right panel of fig (\ref{fig:wtot}). It is evident from the right panel of fig (\ref{fig:wtot}), that as we increase the value of $a_t$, DE+radiation dominate over the matter sector and we see no era of dust domination which is crucial for structure formation. Thus, we limit $a_t\leq 0.1$. Finally, we compare the expansion rate of PFDE model and $\Lambda$CDM using the best fit model parameters inferred from the combination of all data sets. To appreciate the increase in expansion rate in our model, we present our results normalized with respect to the expansion rate of $\Lambda$CDM. In the left and right panel of \cref{fig:Hcom}, we show the dependence of $H_{PFDE} / H_{\Lambda}$ on $n$ and $a_t$ respectively. In all cases the inferred Hubble parameter in the PFDE model is larger than the $\Lambda$CDM one.

\paragraph*{\textbf{Perturbations}}
The evolution of perturbations depends on the sound horizon of DE. In the FLRW Universe, the sound horizon of DE with effective sound speed $c_s$ is defined as
\begin{eqnarray}
    r_s (\eta) = \int \frac{c_s}{ a \,\mathcal{H}} \,d \eta, \label{eq:rs}
\end{eqnarray}
where $\mathcal{H} (\equiv a' / a)$ is the conformal Hubble constant, and $'$ denotes derivative with respect to conformal time $\eta$. Depending on the properties of the fluid, it will have different clustering effects, and $c_s$ plays crucial role in the evolution of DE perturbation. 
If $c_s$ to close to unity, DE perturbations are suppressed by pressure. As a result DE does not cluster except on scales comparable to the $r_s$. In the case of $c_s \ll 1$, DE starts clustering as a dark matter component with the consequence of affecting matter perturbations. Finally, a more general effective speed of sound $0\leq c_s^2\leq 1$, for example from non-canonical scalar fields or unparticles can result in different forms of clustering. 

\begin{figure}[!ht]
    \centering
    \includegraphics[scale=0.63]{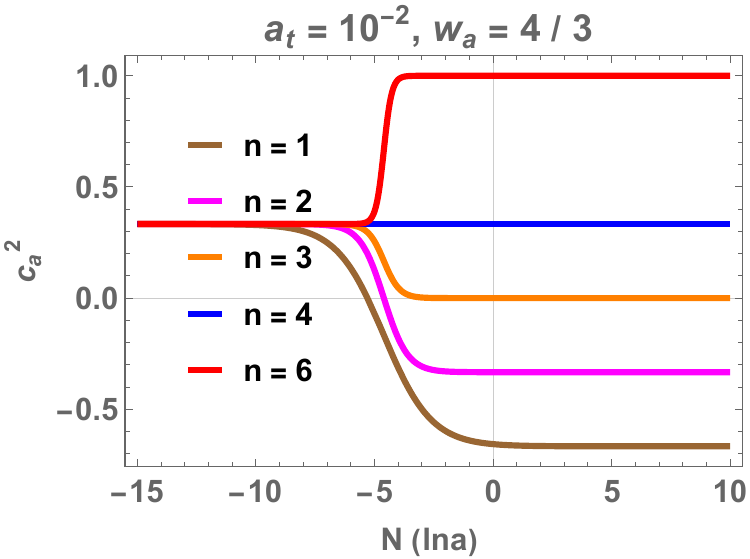}
    \includegraphics[scale=0.63]{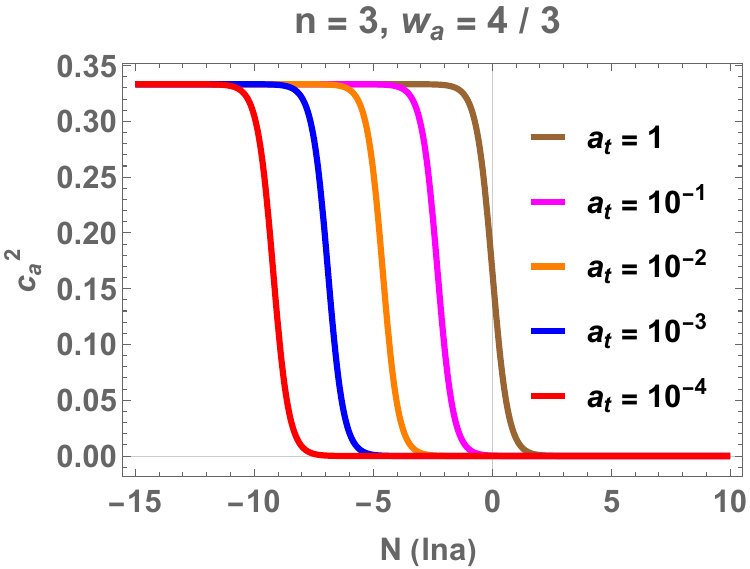}
    \caption{The evolution of adiabatic sound speed $c_a^{2}$ for PFDE model parameters. The left panel shows that adiabatic sound speed becomes negative at the late time for $n < 3$. The right panel illustrates the transition of $c_a^2$ for different values of $a_t$.}
    \label{fig:ca}
\end{figure}

Any phenomenological fluid is characterized by the energy density $\rho_{DE}$, pressure $p_{DE}$, momentum density, and anisotropic stress $\sigma_{DE}$. The pressure   $\delta p_{DE}$ and energy density perturbations $\delta \rho_{DE}$ are related by the effective sound speed which is gauge dependent quantity $c_{s}^{2} = \frac{\delta p_{DE}}{\delta \rho_{DE}}$. 
For canonical scalar field models, it is automatically set to unity. To avoid such gauge ambiguities, we consider the gauge invariant formulation of pressure perturbation Fourier space allowing for a general effective sound speed discussed in \cite{Ma:1995ey}.
\begin{eqnarray}
    \delta p_{DE} & = & c_s^{2}\, \delta \rho_{DE} + 3 \, \mathcal{H}\,\left(c_{s}^{2} - c_{a}^{2} \right)\,\left(1  + w_{DE} \right) \rho_{DE} \, \theta_{DE} / k^{2} \,.
\end{eqnarray}
Here $\theta_{DE}$ is the velocity divergence, and $c_{a}^{2}$ is the adiabatic sound speed for our phenomenological fluid, which takes the following form
\begin{eqnarray}
    c_{a}^{2} \equiv w_{DE} - \frac{w_{DE}'}{3\,\mathcal{H}\,\left(1 + w_{DE}\right)} = w_{DE} - \frac{n}{3\,w_a} \,\left(1 + w_{DE} - w_{a}\right), \label{eq:ca2}
\end{eqnarray}
where for $w_a=4/3$ we have $c_{a}^{2} = w_{DE}(1 - \frac{n}{4}) + \frac{n}{12}$. The adiabatic sound speed and its dependence on the model parameters are depicted in figure \ref{fig:ca}. Notice that for $n<3$ the adiabatic speed of sound tends to negative values, raising the danger of instabilities. We will see that even without addressing this theoretical issue, the data will prefer $n\geq3$.

We write the perturbation equations in synchronous gauge with the following line element:
\begin{eqnarray}
    d\,s^{2} = a^{2}\left(\eta \right) \,\left[ - d\,\eta^{2} + \left( \eta_{i j} + h_{ij}\right)\,d x^i d x^j\right],
\end{eqnarray}
where $h ( \equiv h_{i}^{i})$ is the metric perturbation and $\sigma_{DE}$ is the anisotropic stress \cite{PhysRevD.22.1882}. Within the gauge invariant formalism, we express the perturbations as equations of the density contrast $\delta_{DE}$ and the velocity divergence $\theta_{DE}$:
\begin{eqnarray}
    \delta'_{DE} &=& -\left(1 + w_{DE}\right)\,\left(\theta_{DE} - 3\,h'\right) - 3\mathcal{H}\,\left( c_s^{2} - w_{DE} \right)\delta_{DE}\,, \\ 
   \theta_{DE}' &=& -\mathcal{H}\,\left(1 - 3 w_{DE}\right)\,\theta_{DE} - \frac{w_{DE}'}{1 + w_{DE}} \,\theta_{DE} + \frac{c_s^{2}}{1 + w_{DE}} \, k^{2}\,\delta_{DE} + k^{2}\left( h - \sigma\right).
\end{eqnarray}

It is straightforward to figure out that perturbations of dark energy for a phenomenological fluid offers two extra parameters of the model compared to canonical scalar field, namely, the effective sound speed $c_s^{2}$ and the anisotropic shear/stress $\sigma_{DE}$. In this work, we shall always consider a perfect fluid, so $\sigma_{DE}\equiv 0$. For the effective sound speed $c_s$ there are various possibilities. The effective sound speed $c_s$ importance cannot be underestimated as it reveals the microphysics associated to dark energy. For a barotropic fluid it is equivalent to adiabatic sound speed of the fluid,  $c_s\equiv c_{a}$. As such no free parameters are introduced. A nice example of such a scenario is the Unparticles Dark Energy (UDE) which offers a possible resolution of Hubble and LSS tension simultaneously \cite{Ben-Dayan:2023rgt,Artymowski:2020zwy,Artymowski:2021fkw}.
Attempting to study different microphysics, we shall consider the following scenarios:

\begin{enumerate}
\item Allowing dark energy either to be relativistic $c_s =1$ or be non-relativistic $c_s = 0$.
 
    \item  A perfect fluid, with the $c_s$ as a free parameter, $0\leq c_s\leq 1$.
\end{enumerate}
\section{Data Sets and Methodology} \label{sec:data} 
In our analysis, we use the following publicly available data sets:   

\begin{itemize}
\item{ \textbf{Planck 2018 CMB :}} We utilize the \texttt{Planck 2018} likelihood for the CMB data, which consists of the low-$\ell$ TT, low-$\ell$ EE, and high-$\ell$ TTEETE power spectra \cite{Planck:2019nip}. We also use the \texttt{Planck 2018} lensing likelihood \cite{Planck:2018lbu}, which has an important role in the LSS analysis of the late Universe. 

\item{\textbf{Baryon Acoustic Oscillations (BAO) and RSD measurements:}} We use the measurements from the SDSS DR7 Main Galaxy Sample (MGS) \cite{Ross:2014qpa} and 6dF galaxy survey \cite{Beutler:2011hx} measurements at $z = 0.15$ and $z = 0.106$ respectively. In addition to that, we also include BAO and $f\,\sigma_8$ measurements (where f is the linear growth rate) from BOSS DR12 $\&$ 16 at $z = 0.38,0.51,0.68$ \cite{BOSS:2016wmc,Bautista:2020ahg,Gil-Marin:2020bct,eBOSS:2020yzd}, QSO measurements at $z = 1.48$ \cite{Neveux:2020voa,Hou:2020rse} and Ly-$\alpha$ auto-correlation and cross-correlation with QSO at $z = 2.2334$ \cite{duMasdesBourboux:2020pck}. 

\item{\textbf{Large Scale Structure:}}
\begin{itemize}
    \item \textbf{DES}: Dark Energy Survey includes measurements from shear-shear, galaxy-galaxy, and galaxy-shear two-point correlation functions, referred to as "$3 \times 2$ pt", measured from 26 million source galaxies in four redshifts bins and 650,000 luminous red lens galaxies in five redshifts bins, for the shear and galaxy correlation functions \cite{DES:2021wwk}. DES $3\times2$ pt likelihood gives $S_8 = 0.773^{+0.026}_{-0.020}$ and $\Omega_m = 0.267^{+0.030}_{-0.017}$  for $\Lambda$CDM model. To avoid the computational expenses, we use the Gaussian prior on $S_8$ which effectively summarize the DES likelihood. We also conform that using the Gaussian prior and DES likelihood provide a similar constraint.
    \item \textbf{Weak Lensing Measurements:} In addition to the DES measurements, we also use the measurements from KiDS+VIKING-450 and Subaru Hyper Suprime-Cam (HSC) providing constraints on $S_8$ and $\Omega_m$. In this case also we use Gaussian priors to include their effects. We use $S_8 = 0.737^{+0.040}_{-0.036} $ and $S_8 =  0.780^{+0.030}_{-0.033}$ for KiDS and HSC measurements respectively.
\end{itemize}

\item{\textbf{Supernovae Pantheon:}} The Pantheon data set is a collection of the absolute magnitude of 1048 supernovae distributed in redshift interval $0.01 < z < 2.26 $ \cite{Pan-STARRS1:2017jku}. Many times we will simply refer to this data set as SN.

\item{\textbf{$H_0$ from SH0ES:}} We use latest local measurement of $H_0 = 73.04 \pm 1.04\, \text{km/sec/Mpc}$ from the \texttt{SH0ES} team \cite{Riess:2021jrx}. Many times we will simply refer to this data set as $H_0$.
\end{itemize}

We consider several combinations of datasets to assess the parameter constraints of our phenomenological fluid dark energy model. Our aim is to compare the value of $H_0$ and $S_8$ inferred considering our model compared to the baseline model $\Lambda$CDM. In order to quantify the degree of tension between the different estimates of $H_0$, we adopt the following measure to evaluate the improvement of our model compared to $\Lambda$CDM. We express the tension in terms of standard deviations $\sigma$ for $H_0$ and $S_8$ as 
\begin{eqnarray}
    \# \,\sigma_{H_0} &=& \left| \frac{H_0^{M} - H_0^{\text{SH0ES}}}{\sqrt{\sigma^2_{H_0^{M}} +\sigma^2_{H_0^{\text{SH0ES}}}}} \right| \,,\\
    \# \,\sigma_{S_8} &=& \left| \frac{S_8^{M} - S_8^{\text{LSS}}}{\sqrt{\sigma^2_{S_8^{M}} + \sigma^2_{S_8^{\text{LSS}}}}} \right| \,.
\end{eqnarray}
Here $x^{M}$ and $\sigma_x^{M}$ are the mean value of parameter $x$ and its variance in a given model respectively. In the case of two measurements with asymmetric error bars $X^{\sigma_{X,\text{up}}}_{\sigma_{X,\text{down}}},Y^{\sigma_{Y,\text{up}}}_{\sigma_{Y,\text{down}}}$, then  the tension $\sigma$ between the two measurements is:
\begin{equation}
    \sigma = \begin{cases} 
\frac{X - Y}{\sqrt{\sigma_{X,\text{down}}^2 + \sigma_{Y,\text{up}}^2}} & \text{if } X > Y \\
\frac{Y - X}{\sqrt{\sigma_{X,\text{up}}^2 + \sigma_{Y,\text{down}}^2}} & \text{if } X \leq Y.
\end{cases}
\end{equation} 
We investigate the following combination of datasets: 
\begin{enumerate}
    \item Planck 2018 CMB TTTEEE power spectrum data which is one of the sources of present cosmic tensions. For simplicity, we call this dataset \textbf{Planck 2018 CMB}.
    
    \item  Combination of Primary Planck 2018 TTTEEE+ lensing data with BAO, SNe, and $H_0$ priors. This combination will help us understand the impact of adding other non-CMB datasets on the proposed model. We represent this combination as \textbf{CBSH}.

    \item To understand the large-scale structure ($S_8$) tension, we also consider the full LSS data including DES-Y1, HSC, and KIDS along with Primary Planck 2018 CMB including lensing power spectra, BAO, and SNe datasets with SHOES prior. We denote this combination as \textbf{CBSHDK} \footnote{To analyze all LSS experiments as Gaussian prior on $S_8$ for each data, there exists a possibility of double counting the LSS information and biasing the results as pointed out in \cite{Kilo-DegreeSurvey:2023gfr}. However, to circumvent this issue, we also perform the MCMC analysis using only a single prior on $S_8$. The results are tabulated in \cref{apn:cs1,apn:cs0,apn:csopen}. We thank Gen Ye for pointing out this issue.}.

    \item Finally, we remove the $H_0$ while using all other datasets used previously and refer to it as \textbf{CBSDK}.

\end{enumerate}
The rationale for the different combinations is to see the effect of each data set on the inferred parameters, and allow for amelioration of the tension. For example, if \textbf{CBSH} prefers a lower value of $S_8$, closer to the WL values, and with significant $\Delta \chi^2$ improvement, then the tension is ameliorated compared to $\Lambda$CDM.
The models we consider are the $\Lambda$CDM as baseline model, and extensions according to various possible PFDE.
$\Lambda$CDM model has the usual $6$ free independent parameters: the Baryon $\Omega_b\,h^2$ and Cold Dark Matter $\Omega_c\,h^2$ relative energy densities, the Hubble parameter $H_0$ or the angular scale $\theta_s$, the amplitude $A_s$ and tilt $n_s$ of the primordial power spectrum and $\tau_{reio}$ quantifying optical depth to reionization. The PFDE model offers several extra parameters namely : scale factor $a_t$ at which dark energy eos switches sign, $n$ the sharpness of transition which in turn affects the adiabatic sound speed $c_a^2$ as shown in figure \ref{fig:ca}, and finally the effective sound speed of perturbations $c_s^{2}$. The analysis presented in this paper assumes the models with parameters $a_t$, $n$ and different values of $c_s^{2}$ as
\begin{itemize}
    \item \textbf{Canonical Emergent Dark Energy :} To begin with, we examine the specific case of setting $c_{s}^2 = 1$. These models fall into the category of "Canonical Emergent Dark Energy," which can be realized by employing a canonical scalar field with a suitable potential. By assigning this designation, we distinguish them as a distinct subclass within the broader framework of emergent dark energy models.\cite{Barreiro:1999zs,Chiba:2012cb,Caldwell:2003vp,Caldwell:2000wt,Caldwell:2003hz,Guarnizo:2020pkj,Mehrabi:2015lfa,Gomez:2020sfz}  
    \item \textbf{Clustering Emergent Dark Energy :} Next, we assign a value of  $c_{s}^2$ to zero. Under this circumstance, perturbations in DE exhibit a non-relativistic behavior and cluster akin to dark matter. Numerous investigations have focused on understanding the evolution of these perturbations and the process of structure formation in the context of clustering dark energy.\cite{Batista:2021uhb,Dakin:2019vnj,Hassani:2019lmy,Hassani:2020buk,Mota:2004pa,Bertacca:2008uf}
    \item \textbf{Non-Canonical Emergent Dark Energy :} Moreover, we extend the parameter $c_{s}^{2}$ to span the range from 0 to 1. This class of models, known as non-canonical emergent dark energy models, arises from nonstandard scalar field models. In these models, the properties of dark energy are described by considering alternative formulations of scalar fields, leading to emergent behavior that deviates from canonical expectations.\cite{Armendariz-Picon:2000ulo,Joseph:2021omr,DeDeo:2003te,Erickson:2001bq,Luongo:2018lgy,Luongo:2018lgy,Luongo:2014nld}
\end{itemize}

We sample the posterior distributions of the parameters describing the aforementioned models by using the Markov Chain Monte Carlo (MCMC) method. The chains are produced using the cosmological MCMC sampler \texttt{Cobaya} \cite{Torrado:2020dgo} in conjunction with modified publicly available Einstein-Boltzmann code \texttt{CAMB} \cite{Lewis:1999bs}. The convergence of chains are guaranteed by the Gelman-Rubin parameter \cite{Gelman:1992zz} with $R-1 < 0.03$. We constrain the standard cosmological parameters for all cosmologies with with uniform priors - the baryon matter density $\Omega_b h^{2} \in [0.005, 0.1]$, the cold dark matter density $\Omega_{c}h^{2} \in [0.001,0.99]$, the amplitude of primordial curvature spectrum amplitude $ {\rm{ln}}(10^{10} A_s) \in [1.6,3.9]$ evaluated at suitable pivot scale, $k = 0.05\, Mpc^{-1}$ along with its tilt $ n_s \in [0.8,1.2] $, the reionization optical depth $\tau_{reio} \in [0.01,0.8] $ and the present value of Hubble parameter $ H_0 \in [20,100] $. We use the standard three neutrino description with one massive with mass, $m_{\nu}$ = 0.06 eV, and two massless neutrinos. The posterior distributions are in the Supplementary Material. We work with uniform prior for $a_t$, $n$ and $c_s^2$ with prior edges given by $[0.01,0.1]$, $[1,6]$ and $[0,1]$ respectively.

\section{Results} \label{sec:results}
In the following, we discuss the results obtained using the methods and datasets described in the previous section. This section covers the impact of emergent dark energy cosmologies on $H_0$ and $S_8$ tensions. We present the 68 \% CL parameter inferences for concordance $\Lambda$CDM model in \cref{tab:LambdaCDM}. We will compare the results of different models to the $\Lambda$CDM parameter constraints.

\begin{table}[!ht]
    \centering
    \textbf{The mean $\pm 1 \sigma$ constraints for $\Lambda$CDM model}
    \vspace{1 em}  \\
    \scalebox{1}{
    \begin{tabular}{|c|c|c|c|c|}
    \hline
    Parameter & Planck 2018 CMB & CBSH & CBSHDK & CBSDK \\
    \hline
    $H_0$ & $67.30\pm 0.65$ & $68.09^{+0.63}_{-0.41}$ & $68.13^{+0.86}_{-0.32}$ & $68.16^{+0.37}_{-0.32}$ \\
    $\sigma_8$ & $0.8117\pm 0.0079$ & $0.8102\pm 0.0063$ & $0.8055\pm 0.0058$ & $0.8056\pm 0.0060$ \\
    $S_8$ & $0.833\pm 0.017$ & $0.818^{+0.010}_{-0.015}$ & $0.8119^{+0.0082}_{-0.017}$ & $0.8114\pm 0.0094$ \\
    $\Omega_m$ & $0.3162\pm 0.0089$ & $0.3056^{+0.0051}_{-0.0083}$ & $0.3048^{+0.0039}_{-0.011}$ & $0.3043^{+0.0040}_{-0.0048}$ \\
    $ \# \sigma_{H_0}$ & $4.68$ & $4.07$ & $3.63$ & $4.42$ \\
    $ \# \sigma_{S_8}$ & $2.28$ & $1.89$ & $1.72$ & $1.81$ \\
    Total $\chi^2$ & $2764.17$ & $3854.028$ & $3859.8118$ & $3846.0773$ \\
    $\Delta \chi^2$ & $-$ & $-$ & $-$ & $-$ \\
    \hline
    \end{tabular}
    }
    \caption{ The mean $\pm 1 \sigma$ constraints on the cosmological parameters inferred from the various datasets for $\Lambda$CDM model.}
    \label{tab:LambdaCDM}
\end{table}

\subsection{Canonical Emergent Dark Energy Cosmology}

We begin our exploration by delving into the canonical emergent dark model with $c_s^2 = 1$, laying the foundation for our subsequent analysis. In order to deduce the constraints on the model parameters, we employ various combinations of data sets as outlined in section \ref{sec:data}. The resulting parameter constraints at the $68\%$ confidence level (CL) are presented \cref{tab:cs2equals1}.

Considering Planck 2018 CMB data, we observe a slight increase in the derived value of $H_0 = 68.27 \pm 0.75$ for the canonical emergent dark energy models, compared to the value of $H_0 = 67.30 \pm 0.65$ for $\Lambda$CDM. This increase is consistent across other combinations of data sets, resulting in a range of $H_0$ values spanning from 68.3 to 69.03 km/sec/Mpc in all the considered analyses. Consequently, the discrepancy known as the Hubble tension diminishes to a range of  $\sim 3 \sigma$ confidence level.

\begin{figure}[!ht]
    \centering
    \hspace*{-1.9cm}
    \includegraphics[scale = 0.8]{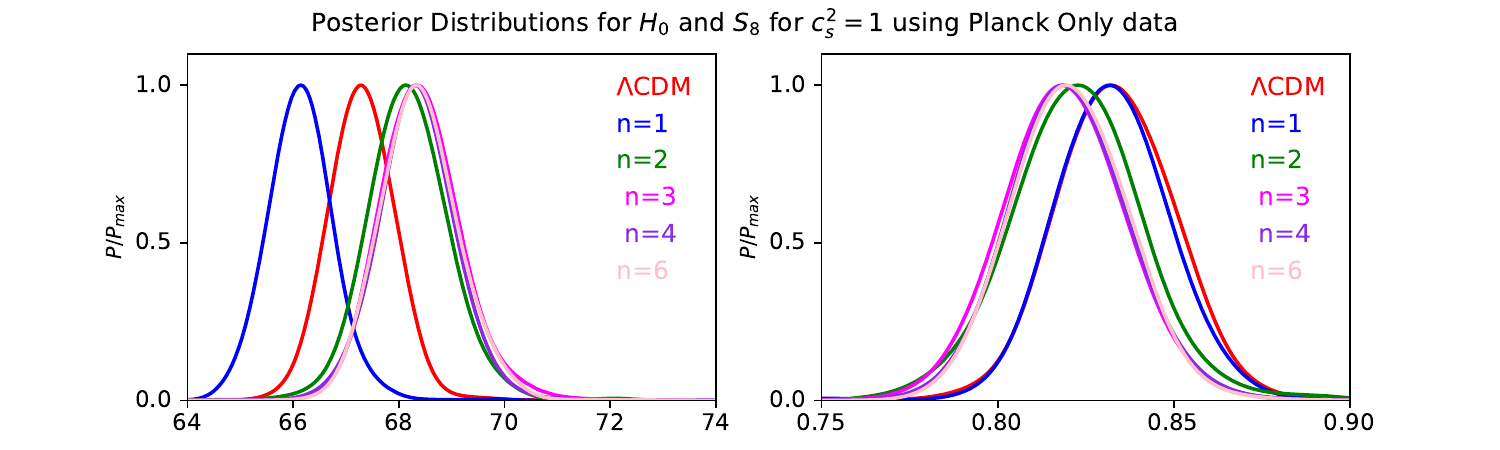}
    \caption{Normalized posterior distributions for $H_0$ (in km/sec/Mpc) and $S_8$ in left and right panels respectively for different choices of $n$ for Emergent Dark Energy model with $c_s^2 = 1$ using  Primary Planck 2018 CMB data. We also show the constraints for $\Lambda$CDM model for comparison. }
    \label{fig:probH0S8cs=1}
\end{figure}

Moreover, we find a marginal decrease in the inferred values of $S_8$ within this framework. For instance, when considering only the Planck data, the inferred value of $S_8$ is $0.82 \pm 0.016$, whereas other data sets yield similar values ranging from $0.82$ to $0.804$. This decrease in $S_8$ can be attributed to the reduction in matter energy density, which exhibits a decrease of approximately 1\%.
 \begin{figure}[!ht]
    \centering
    \hspace*{-1.9cm}
    \includegraphics[scale = 1]{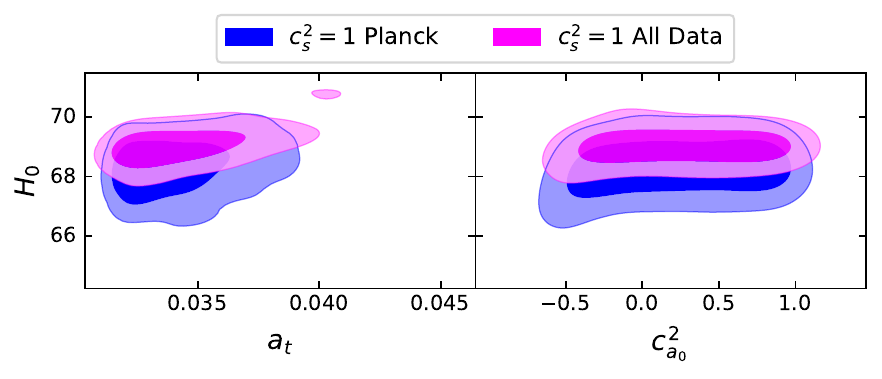}
    \hspace*{-1.9cm}
    \includegraphics[scale = 1]{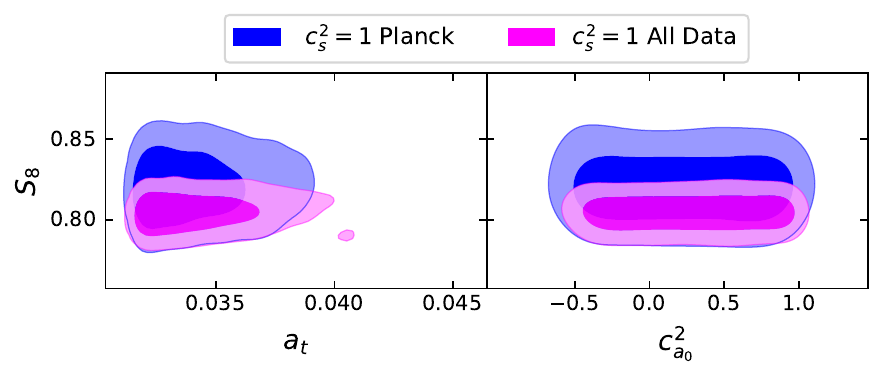}
    \caption{2-dimensional marginalized posterior distributions for model parameters $a_t$ and $c_{a_0}^{2}$ (and hence $n$) with $H_0$ and $S_8 $ in upper and lower panels for canonical emergent dark energy model. The blue and magenta correspond to Primary Planck 2018 CMB data and All Data described in section \ref{sec:data}}
    \label{fig:cs1H0S8}
\end{figure}
Moving forward, we shift our focus to the inference of the model parameters $a_t$ and either $n$ or $c_{a_0}^2$. In order to derive the functional dependence of $c_{a_0}^2$ on other parameters, one can simply evaluate the eq. \ref{eq:ca2} at $z=0$. Our analysis reveals that the $68\%$ confidence evidence points to $10^{2}\,a_t = 3.4 ^{+0.071}_{-0.023}$. Additionally, we constrain the adiabatic sound speed of dark energy at the present time, denoted as $c_{a_0}^{2}$. The analysis indicates positive adiabatic sound speed with $c_{a_0}^{2} = 0.27^{+0.60}_{-0.36}$ at the $68\%$ CL when considering only the Planck data. Furthermore, we establish a lower bound on the parameter $n$, demonstrating that cosmological data favors the canonical emergent dark energy model while excluding the presence of ghost or gradient stability. Our findings indicate $n > 2.88$, and Figure (\ref{fig:ca}) confirms that for $n > 2$, the adiabatic sound speed lies within the range of $[0, 1]$.
\begin{table}[!ht]
    \centering
    \textbf{The mean $\pm 1 \sigma$ constraints for canonical emergent dark energy model ($c_s^{2} = 1$) model}
    \vspace{1 em}  \\
    \scalebox{1}{
    \begin{tabular}{|c|c|c|c|c|}
    \hline
    Parameter & Planck 2018 CMB & CBSH & CBSHDK & CBSDK \\
    \hline
    $H_0$ & $68.27\pm 0.75$ & $68.94^{+0.43}_{-0.62}$ & $69.03\pm 0.50$ & $68.72\pm 0.39$ \\
    $\sigma_8$ & $0.8112\pm 0.0087$ & $0.8134^{+0.0067}_{-0.0078}$ & $0.8078\pm 0.0066$ & $0.8070\pm 0.0062$ \\
    $S_8$ & $0.820\pm 0.016$ & $0.812\pm 0.011$ & $0.8043\pm 0.0092$ & $0.8081\pm 0.0087$ \\
    $\Omega_m$ & $0.3068\pm 0.0092$ & $0.2993^{+0.0067}_{-0.0054}$ & $0.2974\pm 0.0054$ & $0.3009\pm 0.0046$ \\
    $10^{2}\,a_t$ & $3.400^{+0.071}_{-0.023}$ & $3.50^{+0.11}_{-0.32}$ & $3.425^{+0.080}_{-0.26}$ & $3.368^{+0.056}_{-0.20}$ \\
    $c_s^{2}$ & $1$ & $1$ & $1$ & $1$ \\
    $n$ & $> 2.86$ & $> 3.08$ & $> 3.02$ & $> 2.88$ \\
    $c_{a_0}^2$ & $0.23\pm 0.46$ & $0.28\pm 0.41$ & $0.27\pm 0.43$ & $0.24\pm 0.44$ \\
    $ \# \sigma_{H_0}$ & $3.72$ & $3.64$ & $3.47$ & $3.88$ \\
    $ \# \sigma_{S_8}$ & $1.92$ & $1.80$ & $1.63$ & $1.73$ \\
    Total $\chi^2$ & $2766.1977$ & $3852.5094$ & $3857.234$ & $3846.1532$ \\
    $\Delta \chi^2$ & $2.0277$ & $-1.5186$ & $-2.5778$ & $0.0759$ \\
    \hline
    \end{tabular}
    }
    \caption{The mean $\pm 1 \sigma$ constraints on the cosmological parameters inferred from the various datasets for canonical emergent dark energy model ($c_s^{2} = 1$) model.}
    \label{tab:cs2equals1}
\end{table}

We evaluate the goodness-of-fit for various combinations in the canonical emergent dark energy scenario and compare them with $\Lambda$CDM. We observe that the fit to the Planck-only data yields a larger value of $\Delta \chi^2 = 2.0277$, indicating a less favorable fit. However, as we combine the Planck data with several other data sets, we witness an improvement reaching approximately $\Delta \chi^2=-2.6$ in our analysis.

\begin{figure}[!ht]
    \hspace*{-1.9cm}
    \includegraphics[scale = 0.8]{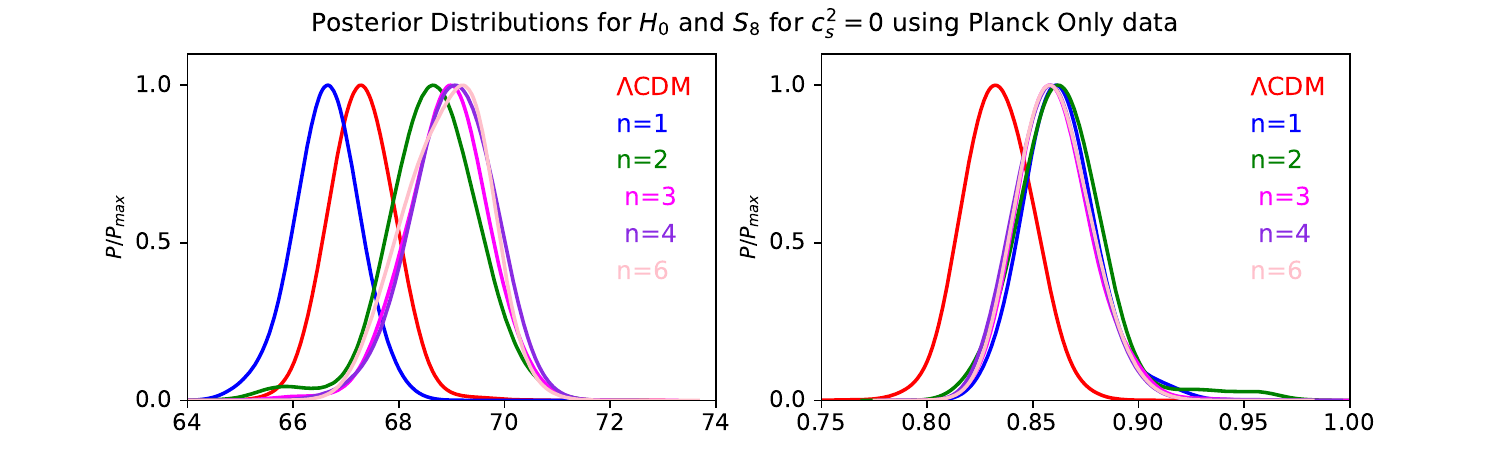}
    \caption{Normalised posterior distributions for $H_0$ (in km/sec/Mpc) and $S_8$ in left and right panels respectively for different choices of $n$ for the Clustering Emergent Dark Energy model with $c_s^2 = 0$ using  Primary Planck 2018 CMB data. We also show the constraints for the $\Lambda$CDM model for comparison. }
    \label{fig:probH0S8cs=0}
\end{figure}

We also study the impact of different values of $n$ which in turn means that $c_{a_0}^{2}$ is fixed. We fixed the values of $n $ spanning from 1 to 6. The resulting constraints for different combination are shown in Appendix  \ref{apn:cs1} (\cref{tab:Planckcs1,tab:CBSHcs1,tab:CBSHDcs1,tab:CBSHDKcs1,tab:CBSDKcs1}). The normalized 1-dimensional posterior distributions for $H_0$ in units of km/sec/Mpc and $S_8$ for different values of $n$ in this scenario using the Primary Planck 2018 CMB data in \cref{fig:probH0S8cs=1}. We notice the increase in inferred value of $H_0$ and the decrease in the inferred value of $S_8$ except for $n=1$, which is prone to ghost instability.
This trend is present in other data-sets combinations as well. 

Finally, we close this subsection by discussing the correlation of model parameters, $a_t$ and $c_{a_0}^{2} $ or $n$, with $H_0$ and $S_8$ considering the various datasets. Upper and lower panel of \cref{fig:cs1H0S8} shows the 2-dimensional contours of $H_0-(a_t,c_{a_0}^2)$ and $S_8-(a_t,c_{a_0}^2)$ respectively. The blue contours are using only Planck data, while the magenta contours take into account all data. In all cases we considered $c_s^2=1$. It is evident from the \cref{fig:cs1H0S8} that in canonical emergent dark energy scenarios $H_0$ and $S_8$ are rather insensitive to $n$, and that $a_t$ is constrained. Comparing to $\Lambda$CDM the model infers a higher value for $H_0$ and a lower one for $S_8$. 

 \begin{figure}[!ht]  
    \centering
    \hspace*{-1.9cm}
    \includegraphics[scale = 1]{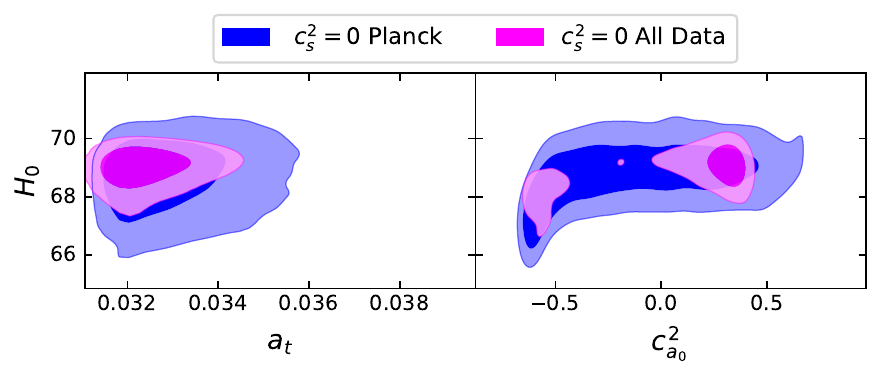}
    \hspace*{-1.9cm}
    \includegraphics[scale = 1]{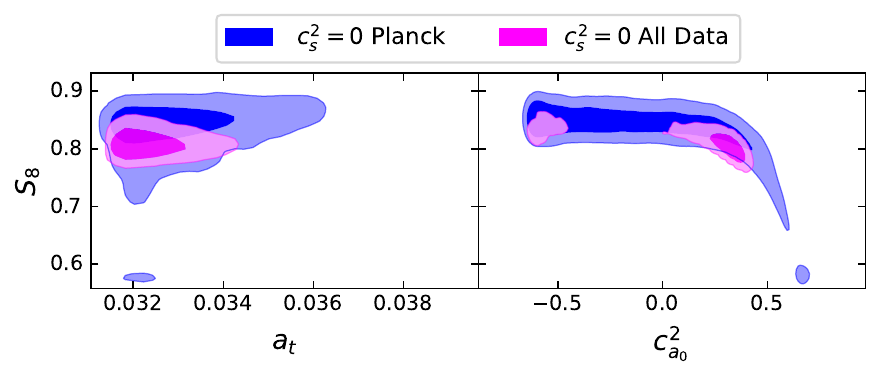}
    \caption{2-dimensional marginalized posterior distributions for model parameters $a_t$ and $c_{a_0}^{2}$ (and hence $n$) with $H_0$ and $S_8 $ in upper and lower panels for clustering emergent dark energy model. These plots show the correlation between model parameters with $H_0$ and $S_8$. The blue and magenta correspond to Primary Planck 2018 CMB data and All data sets described in section \ref{sec:data}}
    \label{fig:cs0H0S8}
\end{figure}

\subsection{Clustering Emergent Dark Energy Cosmology} In this section, we explore the impact of clustering Emergent Dark Energy Cosmologies, which corresponds to $c_s=0$ on various data sets. The parameter constraints at the $68\%$ confidence level (CL) are presented in the \cref{tab:cs2equals0}, alongside the best-fit $\chi^2$ and $\Delta \,\chi^2$ assuming $\Lambda$CDM.

We observe an increase in the derived values of $H_0$ and $S_8$ compared to the Primary Planck 2018 CMB data analysis. Hence, while there is a marginal improvement in the $H_0$ value, the $S_8$ value worsens in comparison to the canonical emergent dark energy cosmologies using Primary Planck 2018 CMB data. 
Because in this model DE can cluster, we find an increment in $\sigma_8 = 0.838^{+0.033}_{-0.0057}$. 
This increase in $\sigma_8$ is solely responsible for the worsening of the $S_8$ tension. The model parameters $10^{2}\,a_t$ and $n$ (or $c_{a_0}^2$) are reported as $3.288^{+0.033}_{-0.12}$ and $<3.17$ ($-0.14^{+0.38}_{-0.50}$), respectively.

Next, we examine the consequences of considering other data sets within this scenario. The present values of the Hubble parameter $H_0$ and the amplitude of matter fluctuations $S_8$ follow the same trends across different data sets and their combinations. The inferred values of $H_0$ and $S_8$ lie within the ranges of $[68 - 69]$ km/sec/Mpc and $[0.843 - 0.808]$, respectively, for the other combinations. The predictions of the model, represented by $a_t$ and $n$ (or the derived parameter $c_{a_0}^2$), reveal a transition redshift $a_t$ of approximately $0.034$. However, in the case of the combination data sets, we find a lower bound on $n < 3.0$, indicating the presence of gradient or ghost instabilities in clustering dark energy. The possible correlations between additional model parameters and $H_0$ and $S_8$ are illustrated in \cref{fig:cs0H0S8}. The 68 $\%$ constraint for different $n$ are shown in Appendix  \ref{apn:cs0}, \cref{tab:Planckcs0,tab:CBSHcs0,tab:CBSHDcs0,tab:CBSHDKcs0,tab:CBSDKcs0}.

Finally, we conclude the discussion on clustering in emergent dark energy by commenting on the $\chi^2$ and $\Delta \chi^2$. In this scenario, we observe a worse fit to the data compared to our baseline model $\Lambda$CDM and the canonical emergent dark energy case. As mentioned earlier, while it is possible to alleviate the Hubble tension in this scenario, the tension related to large-scale structure becomes more pronounced. 
\begin{table}[!ht]
    \centering
    \textbf{The mean $\pm 1 \sigma$ constraints for clustering emergent dark energy ($c_s^{2} = 0$) model}
    \vspace{1 em}  \\
    \scalebox{1}{
    \begin{tabular}{|c|c|c|c|c|}
    \hline
    Parameter & Planck 2018 CMB & CBSH & CBSHDK & CBSDK \\
    \hline
    $H_0$ & $68.7^{+1.0}_{-0.78}$ & $68.61^{+0.67}_{-0.41}$ & $69.00^{+0.53}_{-0.34}$ & $ 68.96\pm 0.40$ \\
    $\sigma_8$ & $0.838^{+0.023}_{-0.0057}$ & $0.846^{+0.012}_{-0.0093}$ & $0.814^{+0.014}_{-0.016}$ & $ 0.8422^{+0.0052}_{-0.0070}$ \\
    $S_8$ & $0.837^{+0.030}_{-0.011}$ & $0.846^{+0.014}_{-0.012}$ & $0.808^{+0.013}_{-0.018}$ & $0.8369^{+0.0084}_{-0.096}$ \\
    $\Omega_m$ & $0.2993^{+0.0093}_{-0.011}$ & $0.3001^{+0.0049}_{-0.0068}$ & $0.2958^{+0.0042}_{-0.0057}$ & $0.2963^{+0.0044}_{-0.0051}$ \\
    $10^{2}\,a_t$ & $3.288^{+0.033}_{-0.12}$ & $3.255^{+0.022}_{-0.092}$ & $3.239^{+0.017}_{-0.077}$ & $3.400^{+0.069}_{-0.23}$ \\
    $c_s^{2}$ & $0$ & $0$ & $0$ & $0$ \\
    $n$ & $< 3.17$ & $< 3.14$ & $3.67^{+0.47}_{+0.12}$ & $> 3.07$ \\
    $c_{a_0}^2$ & $-0.14^{+0.39}_{-0.50}$ & $-0.16\pm 0.30$ & $0.225^{+0.16}_{+0.039}$ & $0.28^{+0.64}_{-0.32}$ \\
    $ \# \sigma_{H_0}$ & $3.00$ & $3.58$ & $3.46$ & $3.66$ \\
    $ \# \sigma_{S_8}$ & $2.41$ & $2.61$ & $1.61$ & $0.96$ \\
    Total $\chi^2$ & $2767.1379$ & $3852.2095$ & $3859.1056$ & $3857.8257$ \\
    $\Delta \chi^2$ & $2.9679$ & $-1.8185$ & $-0.7062$ & $+ 11.74$ \\
    \hline
    \end{tabular}
    }
    \caption{The mean $\pm 1 \sigma$ constraints on the cosmological parameters inferred from the various datasets for clustering emergent dark energy ($c_s^{2} = 0$) model}
    \label{tab:cs2equals0}
\end{table}

 \begin{figure}
    \centering
    \hspace*{-1.9cm}
    \includegraphics[scale = 1.1]{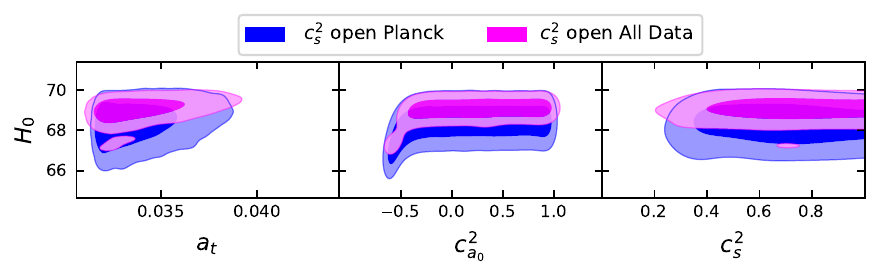}
    \hspace*{-1.9cm}
    \includegraphics[scale = 1.1]{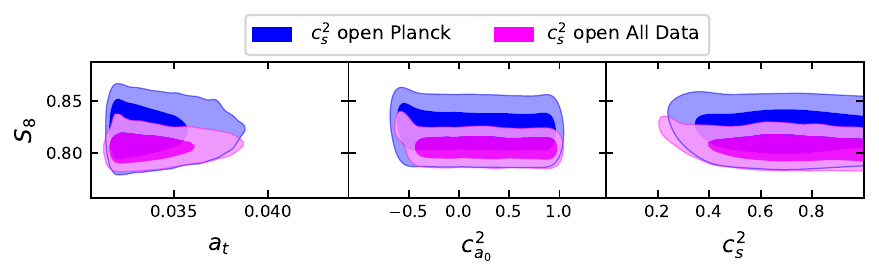}
    \caption{2-dimensional marginalized posterior distributions for model parameters $a_t$ and $c_{a_0}^{2}$ (and hence $n$) with $H_0$ and $S_8 $ in upper and lower panels for non-canonical emergent dark energy model. These plots show the correlation between model parameters with $H_0$ and $S_8$. The blue and magenta correspond to Primary Planck 2018 CMB data and All data sets described in section \ref{sec:data}}
    \label{fig:csopenH0S8}
\end{figure}

\begin{figure}[!ht]
    \hspace*{-1.9cm}
    \includegraphics[scale = 0.8]{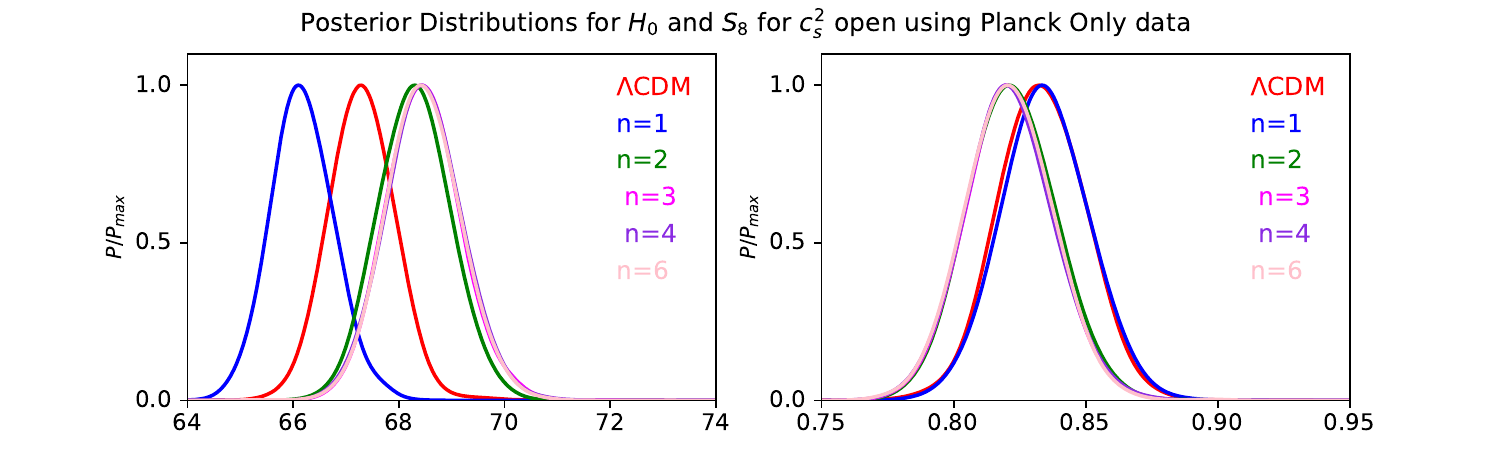}
    \caption{Normalized posterior distributions for $H_0$ (in km/sec/Mpc) and $S_8$ in left and right panels respectively for different choices of $n$ for Emergent Dark Energy model with keeping $c_s^2$ open using  Primary Planck 2018 CMB data. We also show the constraints for the $\Lambda$CDM model also for comparison. }
    \label{fig:probH0S8opencs}
\end{figure}

\subsection{Non-canonical Emergent Dark Energy }  

Finally, we investigate the implications of non-canonical emergent dark energy by allowing the effective sound speed ($c_{s}^{2}$) to vary between 0 to 1, along with other model parameters. We present the parameter constraints for this case in \cref{tab:opencs2}.
Similar to previous cases, we find an increase in the value of $H_0$ and a marginal decrease in $S_8$. Referring to the results from the Primary Planck 2018 Cosmic Microwave Background (CMB) analysis, we infer $H_0=68.30_{-0.72}^{+0.51}$ km/sec/Mpc and $S_8=0.822 \pm 0.017$. The amplitude of matter fluctuations $\sigma_8$ and matter density $\Omega_m$ are constrained to be $0.8133 \pm 0.0093$ and $0.3066 ^{+0.0089}_{-0.010}$ respectively. The tension metrics for $H_0$ and $S_8$ show an improvement of approximately 1 $\sigma$ and $0.5 \sigma$ respectively. Moving to other data sets, the improvement in $H_0$ and $S_8$ remains consistent, resulting in a range of $H_0 \approx 68.8$ km/sec/Mpc and $S_8 \approx 0.81$.

In all cases considered, we find the transition scale factor to be approximately $a_t=0.034$, regardless of the data set used. The steepness parameter $n$ is constrained to ensure that $c_a^2$ remains free from gradient/ghost instabilities and subluminal throughout the evolution. Using the Primary Planck 2018 CMB data in combination with Baryon Acoustic Oscillations (BAO), Supernovae (SNe), and for the SH0ES measurement, we find that $n = 3.8^{+1.8}_{-1.1}$ and, consequently, $c_{a_0}^{2} = 0.26^{+0.62}_{-0.43}$. The effective sound speed of perturbations, $c_{s}^{2}$, is determined to be $0.27 \pm 0.46$ for the same data set. The correlation between the model parameters ($a_t$, $c_{a_{0}}^{2}$, and $c_s^{2}$) with $H_0$ and $S_8$ is depicted in \cref{fig:csopenH0S8}, demonstrating that $H_0$ and $S_8$ are insensitive to the values of  $n$, $c_{a_0}^{2}$, and $c_{s}^2$. The non-canonical emergent dark energy model provides a better fit to the data, regardless of the data sets considered. The difference in $\chi^2$ compared to the $\Lambda$CDM model ranges from -3 to -0.20.

\begin{table}[!ht]
    \centering
    \textbf{The mean $\pm 1 \sigma$ constraints for non-canonical emergent dark energy (open $c_s^{2}$) model}
    \vspace{1 em}  \\
    \scalebox{1}{
    \begin{tabular}{|c|c|c|c|c|}
    \hline
    Parameter & Planck 2018 CMB & CBSH & CBSHDK & CBSDK \\
    \hline
    $H_0$ & $68.30^{+0.84}_{-0.72}$ & $68.82^{+0.45}_{-0.39}$ & $69.00^{+0.48}_{-0.32}$ & $68.73\pm 0.41$ \\
    $\sigma_8$ & $0.8133\pm 0.0093$ & $0.8147^{+0.0066}_{-0.0080}$ & $0.8092^{+0.0059}_{-0.0070}$ & $0.8086^{+0.0060}_{-0.0069}$ \\
    $S_8$ & $0.822\pm 0.017$ & $0.8155^{+0.0091}_{-0.011}$ & $0.8063^{+0.0079}_{-0.010}$ & $0.8098\pm 0.0093$ \\
    $\Omega_m$ & $0.3066^{+0.0089}_{-0.010}$ & $0.3006^{+0.0045}_{-0.0052}$ & $0.2979^{+0.0036}_{-0.0055}$ & $0.3009\pm 0.0046$ \\
    $10^{2}\,a_t$ & $3.374^{+0.056}_{-0.21}$ & $3.423^{+0.082}_{-0.25}$ & $3.381^{+0.061}_{-0.21}$ & $3.338^{+0.045}_{-0.17}$ \\
    $c_s^{2}$ & $0.65^{+0.26}_{-0.17}$ & $0.66^{+0.22}_{-0.19}$ & $0.68^{+0.26}_{-0.14}$ & $> 0.597$ \\
    $n$ & $3.45 \pm 1.45$ & $3.8^{+1.8}_{-1.1}$ & $> 2.95$ & $> 3.06$ \\
    $c_{a_0}^2$ & $0.15\pm 0.48$ & $0.26^{+0.62}_{-0.43}$ & $0.24\pm 0.44$ & $0.27\pm 0.46$ \\
    $ \# \sigma_{H_0}$ & $3.54$ & $3.72$ & $3.52$ & $3.85$ \\
    $ \# \sigma_{S_8}$ & $1.95$ & $1.89$ & $1.68$ & $1.77$ \\
    Total $\chi^2$ & $2765.8647$ & $3851.7445$ & $3856.9329$ & $3846.0359$ \\
    $\Delta \chi^2$ & $1.6947$ & $-2.2835$ & $-2.8789$ & $-0.0414$ \\
    \hline
    \end{tabular}
    }
    \caption{The mean $\pm 1 \sigma$ constraints on the cosmological parameters inferred from the various datasets for non-canonical emergent dark energy (open $c_s^{2}$) model.}
    \label{tab:opencs2}
\end{table}

Next, we explore the impact of different values of $n$ in this scenario. \cref{fig:probH0S8opencs} displays the 1-dimensional normalized probability distribution for $H_0$ and $S_8$ for $n$ ranging from 1 to 6, using the Primary Planck 2018 CMB data. For $n = 1$, the model is disfavored as it yields $\Delta \chi^2 \in [2, 13]$ and predicts a lower $H_0$ and slightly higher value of $S_8$. As observed in the case of open $n$, for $n \in [3, 6]$, we find an improvement in both $H_0$ and $S_8$ simultaneously, providing a better fit compared to the $\Lambda$CDM model and cases with $c_s^2 = 1$ or $0$. The 68 $\%$ constraint for different $n$ are shown in Appendix  \ref{apn:csopen} \cref{tab:Planckopencs,tab:CBSHopencs,tab:CBSHDopencs,tab:CBSHDKopencs,tab:CBSDKopencs}.

\begin{figure}
    \centering
    \includegraphics[scale = 0.5]{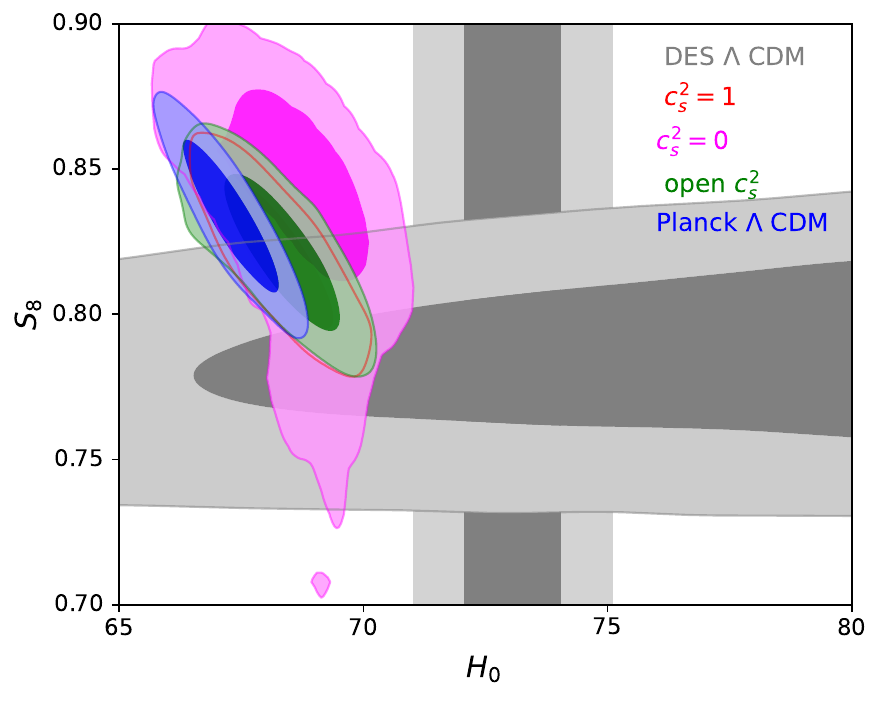}
    \includegraphics[scale = 0.5]{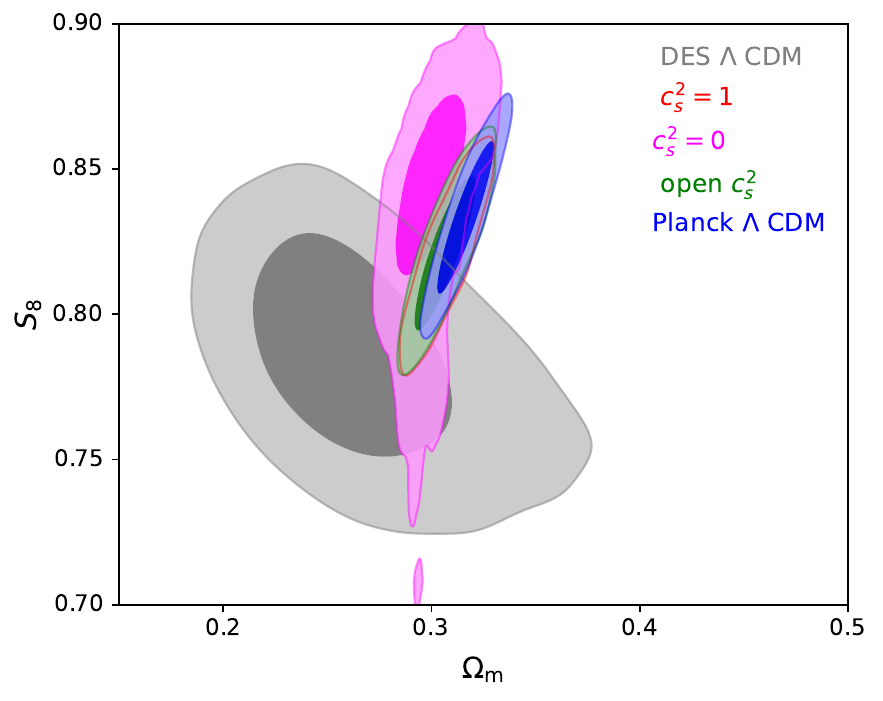}
    \caption{Left Panel: Comparison of $H_0$ and $S_8$ for the $\Lambda$CDM and Emergent Dark Energy with different $c_s^2$. The horizontal and vertical axis represent the $H_0$ in units of km/sec/Mpc  and $S_8$ respectively. We denote as $\Lambda$CDM as blue, $c_s^2 = 1$ as red, $c_s^2 = 0$ as magenta and $c_s^2$ open as green. The dark and light shaded regions correspond to the 68 \% and 95 \% confidence level (CL) respectively. Right Panel: Comparison of $\Omega_m$ and $S_8$ as horizontal and vertical axis respectively with similar color codes. In both plots the local SH0ES and DES measurements are expressed in gray.}
    \label{fig:H0S8}
\end{figure}

We compare all three models analyzed in this paper with $\Lambda$CDM in \cref{fig:H0S8}. The Primary Planck 2018 CMB constraints for $\Lambda$CDM,  and Emergent DE model with $c_s^{2} = 1$, $c_s^{2} = 0$ and $c_s^{2} $ open are denoted as blue red, magenta, and green respectively. The shaded dark and light regions correspond to the $68 \% $ and $95 \% $ confidence levels. respectively. We also compare the results with the latest local SH0ES and DES-Y1 measurements. It is clear from both the left and right panels of \cref{fig:H0S8} that canonical and non-canonical dark energy models offer a marginal increase in values of $H_0$ and a marginal decrease in $S_8$.

\section{Conclusions} \label{sec:conclusions}
We have studied the emergent dark energy model by considering the phenomenological fluid approach. This approach offers to probe the arbitrary sound speed of dark energy perturbations. The emergent dark energy behaves like a cosmological constant ($\Lambda$) at late times and asymptotes to a radiation fluid at early times. As such, it incorporates a tracking mechanism and allows for a rather sudden transition in the DE eos. We employed a phenomenological fluid with 2 parameters to mimic this feature. These two parameters are the scale factor, at which the equation of state of the phenomenological fluid crosses 0, $a_t$, and the steepness parameter $n$. In addition to these parameters, we also have an arbitrary sound speed of dark energy perturbations $c_{s}^{2}$ taking values within the range of 0 to 1. In this paper, we considered the three different case of $c_{s}^2 = 1$, $c_{s}^2 = 0$ and $c_{s}^2 $ open which we assign them as canonical, clustering and non-canonical emergent dark energy models. The consideration of each case is capable of explaining the current accelerated expansion of the Universe. The model further offers a built-in solution to the cosmic coincidence problem. We show the impact of $a_t$ and $n$ on background and perturbation evolution. The background behavior is illustrated in \cref{fig:wde} and \cref{fig:wtot}. We notice that there exist a threshold value of $a_t$ in order to get an era of matter domination for structure formation. The adiabatic sound speed $c_{a}^{2}$ which depends on both parameter $a_t$ and $n$ is shown in \cref{fig:ca}. It is worth mentioning that $c_a^2$ has strong dependence on $n$. Left panel of \cref{fig:ca} clearly shows that for $n \geq 3$, $c_a^2$ lies with in range of $[0,1]$ further constraining the parameter space. 

We then analyze the parameter constraints using the different cosmic data-sets. The aim of this exercise is to check the ability of alleviating the cosmological tensions ($H_0$ and $S_8$) in the emergent dark energy scenario. We consider the various data-sets to assess the impact of model in reducing the cosmic tensions. We notice an overall improvement of both tensions for all models considered. Out of all the model considered in our analysis, we find that keeping $c_s^2$ open i.e. non-canonical emergent dark energy provides the best improvement with respect to $\Lambda$CDM. The canonical model also reduces both cosmic tensions while clustering (non-relativistic) model reduces the Hubble constant tension and worsens the LSS tension. During our analysis, we chose a wide prior on $n \in [1,6]$ , being agnostic about the fact that $n \leq3$ makes the evolution problematic and let the data decide to constrain the value of $n$. Interestingly, we find that most of the data-sets constrain the value of $n$ respecting the theoretical arguments and prefer $n\geq 3$.
The reduction in the Hubble and LSS tension is rather mild and not very significant statistically. This is probably a reflection of the fact that  $H_0$ and $S_8$ seem rather insensitive to $a_t$ and $n$. The main difference in the current analysis compared to the emergent unparticles DE model (UDE) \cite{Ben-Dayan:2023rgt} is that here the adiabatic speed of sound $c_a^2$ and the effective sound speed $c_s^2$ were different, while in the UDE they were equal. Given that in the UDE we got a much more significant reduction of the tensions in a statistically significant way, highlights the important role the adiabatic and effective sound speeds play in inferring the cosmological parameters. 
Given the insensitivity to $n$ it seems that fixing $n$ and $c_s^2$ and leaving only $a_t$ as a free parameter can be competitive with the wCDM model, and can be constrained significantly with current and future data. 
The most intriguing result in our opinion is the fact that the transition redshift is highly limited with $29<1+z<30$. As a result one can think of ways to try and detect such a transition.

\section*{Acknowledgements}
We acknowledge the Ariel HPC Center at Ariel University for providing computing resources that have contributed to the research results reports reported within this paper.
\bibliographystyle{JHEP}
\bibliography{reference}
\appendix
\appendixpage
\renewcommand\thesubsection{\thesection.\arabic{subsection}}

\section{Canonical Emergent Dark Energy} \label{apn:cs1}
\subsection{Constraints for Planck 2018 CMB for different values of n }
    \begin{table}[H]
\begin{center}
 The mean $\pm 1 \sigma$ constraints from Planck 2018 CMB for $c_s^2 = 1$ for different $n$
   \vspace{1 em}  \\
\scalebox{1}{
\begin{tabular}{|c|c|c|c|c|c|}

\hline
\hline

    Params &$n=1$ & $n=2$  & $n=3$ & $n=4$&$n=6$ \\
\hline
\hline
$ H_0 $     & $66.15\pm 0.68 $ & $68.18\pm 0.87 $& $68.42^{+0.68}_{-0.83} $&$68.35\pm 0.75 $ &$68.39^{+0.68}_{-0.79} $ \\
$\sigma_8$  &$0.7973^{+0.0073}_{-0.0084} $  & $0.8122^{+0.0081}_{-0.0097} $ &$0.8115\pm 0.0096 $ & $0.8120\pm 0.0094 $ &$ 0.8128^{+0.0074}_{-0.0090}$ \\
$ S_8$ & $0.832\pm 0.017 $ &$0.823\pm 0.021 $ &$0.819\pm 0.018 $ &$0.820\pm 0.018 $ &$0.821\pm 0.017 $ \\
$ \Omega_m$ & $0.3270\pm 0.0095 $ &$0.308\pm 0.011 $ &$ 0.3055\pm 0.0098$ &$0.3063\pm 0.0094 $ &$ 0.3059\pm 0.0094$ \\
\hline
\hline
$ 10^{2}\,a_t$        & $3.470^{+0.075}_{-0.31} $ & $3.429^{+0.056}_{-0.27} $&$3.421^{+0.057}_{-0.26} $ & $ 3.390^{+0.068}_{-0.23}$&$3.417^{+0.051}_{-0.25} $ \\

\hline 
    \hline
    Total $\chi^2 $ & 2766.0376 &2765.8973& 2765.8248&2766.1938 &2765.7482\\
     $\Delta \chi^2 $ & 1.8676 &1.7273 & 1.6548& 2.0238& 1.5782 \\
\hline
\hline

\end{tabular}
}
\end{center}

 \caption{The mean $\pm 1 \sigma$(best-fit) constraints on the cosmological parameters inferred from the \textit{Planck} 2018 CMB data (including \texttt{TTEETE}) for $\Lambda$CDM and Canonical Emergent DE scenario with different values of $n$ varying from 1 to 6.  }
 \label{tab:Planckcs1}
\end{table}

\subsection{Planck + Lensing + BAO + $H_0$}

\begin{table}[H]
\centering
The mean $\pm 1 \sigma$ constraints from Planck 2018 + Lensing + BAO + SN +$H_0$ 
   \vspace{1 em}  \\

\scalebox{1}{
\begin{tabular}{|c|c|c|c|c|c|}

\hline
\hline

    Parameter         &$n=1$ & $n=2$  & $n=3$ & $n=4$&$n=6$ \\

\hline
\hline
$ H_0 $     & $67.33^{+0.67}_{-0.51} $ & $68.88^{+0.62}_{-0.89}$& $ 68.85^{+0.48}_{-0.59}$&$ 68.96^{+0.49}_{-0.80}$ &$ 68.90^{+0.50}_{-0.60}$ \\
$\sigma_8$  &$ 0.7985^{+0.0065}_{-0.0081}$  & $0.8139^{+0.0065}_{-0.0088} $ &$ 0.8141^{+0.0062}_{-0.0082}$ & $0.8141^{+0.0063}_{-0.0086} $ &$ 0.8137^{+0.0060}_{-0.0081}$ \\
$ S_8$ & $ 0.812^{+0.011}_{-0.015}$ &$ 0.814\pm 0.016$ &$0.8149^{+0.0096}_{-0.012} $ &$ 0.813\pm 0.014$ &$ 0.814^{+0.010}_{-0.011}$ \\
$ \Omega_m$ & $ 0.3105^{+0.0062}_{-0.0092}$ &$0.3000^{+0.0091}_{-0.0081}$ &$ 0.3006\pm 0.0082$ &$ 0.2995^{+0.0084}_{-0.0065}$ &$ 0.3000\pm 0.0075$ \\
\hline
\hline

$ 10^{2}\,a_t$        & $ 3.58^{+0.10}_{-0.41}$ & $3.54^{+0.11}_{-0.37} $&$3.500^{+0.089}_{-0.32} $ & $3.514^{+0.091}_{-0.35} $&$ 3.496^{+0.090}_{-0.32} $ \\

\hline 
    \hline
    Total $\chi^2 $ &3869.1753  & 3852.9131&  3851.876& 3852.3776& 3852.1676 \\
     $\Delta \chi^2 $ &15.5434  & -1.0037&-1.2103 &-0.9626 & -1.2265\\

\hline
\hline

\end{tabular}}
 
 \caption{ The mean $\pm 1 \sigma$(best-fit) constraints on the cosmological parameters inferred from the \textit{Planck} 2018 CMB data (including \texttt{TTEETE} + lensing), BAO, SNe and SH0ES data for $\Lambda$CDM and Canonical Emergent DE scenario with different values of $n$ varying from 1 to 6.  }
 \label{tab:CBSHcs1}
\end{table}

\subsection{Planck + Lensing + BAO + $H_0$ + DES-Y1}

\begin{table}[H]
\centering
 Constraints from Planck 2018 + Lensing + BAO + SN +$H_0$ + DES for Canonical Emergent DE
   \vspace{1 em}  \\

\scalebox{1}{
\begin{tabular}{|c|c|c|c|c|c|c|}

\hline
\hline

    Parameter         &$n=1$ & $n=2$  & $n=3$ & $n=4$&$n=6$ & $n $ open  \\

\hline
\hline
$ H_0 $          & $ 67.59\pm 0.64$ & $ 68.75^{+0.51}_{-0.41}$ & $68.82^{+0.45}_{-0.40} $ & $ 69.01^{+0.44}_{-0.56}$& $  68.99^{+0.45}_{-0.51} $ & $68.93^{+0.53}_{-0.41}$ \\
$\sigma_8$       & $0.7972^{+0.0064}_{-0.0081} $ & $ 0.8110^{+0.0062}_{-0.0074}$ & $0.8117^{+0.0059}_{-0.0073} $ & $ 0.8126^{+0.0061}_{-0.0079}$& $ 0.8123^{+0.0063}_{-0.0075}$ & $0.8107^{+0.0065}_{-0.0075}$\\
$ S_8$       & $ 0.8063^{+0.0095}_{-0.012}$ & $ 0.8121^{+0.0082}_{-0.012}$& $ 0.8123^{+0.0087}_{-0.011}$ & $ 0.810\pm 0.011$& $0.810\pm 0.011 $ & $ 0.8092^{+0.0089}_{-0.010}$\\
$ \Omega_m$       & $ 0.3070\pm 0.0081$ & $ 0.3009^{+0.0046}_{-0.0063}$& $0.3005^{+0.0043}_{-0.0057} $ & $ 0.2984^{+0.0061}_{-0.0053}$& $ 0.2986\pm 0.0066$ & $0.2990^{+0.0046}_{-0.0062}$\\
\hline
\hline

$ 10^{2}\,a_t$              & $ 3.58^{+0.11}_{-0.41}$ & $ 3.456^{+0.096}_{-0.28}$& $3.441^{+0.096}_{-0.27} $ & $ 3.47^{+0.10}_{-0.29}$& $3.47^{+0.10}_{-0.30} $ & $3.462^{+0.094}_{-0.29}$\\
$n$ & 1 & 2 & 3 & 4 & 6 & $> 3.05$ \\

\hline 
    \hline
    Total $\chi^2 $ &3869.5714& 3853.0243& 3852.8177&3853.0654&3852.8015 & 3854.1713\\
     $\Delta \chi^2 $ &13.843  &-2.7041 &-2.9107 &-2.663 &-2.9269 & -1.55\\

\hline
\hline

\end{tabular}}
 
 \caption{ The mean $\pm 1 \sigma$(best-fit) constraints on the cosmological parameters inferred from the \textit{Planck} 2018 CMB data (including \texttt{TTEETE} + lensing), BAO, SNe, SH0ES and DES-Y1 data for $\Lambda$CDM and Canonical Emergent DE scenario with different values of $n$ varying from 1 to 6.}
 \label{tab:CBSHDcs1}
\end{table}

\subsection{Planck + Lensing + BAO + $H_0$ + DES-Y1 + HSC + KIDS}

\begin{table}[H]
Constraints from Planck 2018 + Lensing + BAO + SN +$H_0$ + DES + HSC + KIDS for Canonical Emergent DE
   \vspace{1 em}  \\
\centering
\scalebox{1}{
\begin{tabular}{|c|c|c|c|c|c|}

\hline
\hline

    Parameter         &$n=1$ & $n=2$  & $n=3$ & $n=4$&$n=6$ \\

\hline
\hline
$ H_0 $          & $ 67.61^{+0.46}_{-0.40}$ & $ 69.01\pm 0.68$& $69.16^{+0.42}_{-0.57} $ & $ 69.16^{+0.39}_{-0.44}$& $ 68.99^{+0.49}_{-0.44}$ \\
$\sigma_8$       & $0.7933^{+0.0055}_{-0.0074} $ &$ 0.8076^{+0.0055}_{-0.0069}$ & $ 0.8088^{+0.0055}_{-0.0080}$ & $ 0.8077^{+0.0058}_{-0.0069}$& $ 0.8083\pm 0.0071$ \\
$ S_8$       & $ 0.8016^{+0.0082}_{-0.011}$ & $ 0.8041^{+0.0088}_{-0.0099}$& $ 0.804\pm 0.011$ & $ 0.803\pm 0.010$& $0.8057^{+0.0079}_{-0.0097} $ \\
$ \Omega_m$       & $ 0.3064^{+0.0046}_{-0.0064}$ & $ 0.2975^{+0.0049}_{-0.0057}$& $ 0.2965\pm 0.0078$ & $ 0.2962\pm 0.0049$& $ 0.2981^{+0.0044}_{-0.0057}$ \\
\hline
\hline

$ 10^{2}\,a_t$              & $ 3.529^{+0.089}_{-0.36}$ & $ 3.434^{+0.077}_{-0.26}$ & $ 3.474^{+0.057}_{-0.30}$ & $ 3.441^{+0.091}_{-0.27}$& $ 3.405^{+0.077}_{-0.23}$ \\

\hline 
    \hline
    Total $\chi^2 $ &3873.4982& 3857.2683& 3857.9786& 3856.8311& 3857.0445\\
     $\Delta \chi^2 $ & 13.6864 & -2.5435& -1.8332&-2.9807 & -2.7673\\

\hline
\hline

\end{tabular}}
 
 \caption{ The mean $\pm 1 \sigma$ constraints on the cosmological parameters inferred from the \textit{Planck} 2018 CMB data (including \texttt{TTEETE} + lensing), BAO, SNe, SH0ES, DES-Y1 and weak-lensing data for $\Lambda$CDM and Canonical Emergent DE scenario with different values of $n$ varying from 1 to 6.}
 \label{tab:CBSHDKcs1}
\end{table}

\subsection{ Planck + Lensing + BAO + DES-Y1 + HSC + KIDS}

\begin{table}[H]
Constraints from Planck 2018 + Lensing + BAO + SN + DES + KIDS + HSC for Canonical Emergent DE
   \vspace{1 em}  \\
\centering
\scalebox{1}{
\begin{tabular}{|c|c|c|c|c|c|}

\hline
\hline

    Parameter         &$n=1$ & $n=2$  & $n=3$ & $n=4$&$n=6$ \\

\hline
\hline
$ H_0 $          & $ 67.23^{+0.41}_{-0.32}$ & $ 68.63^{+0.37}_{-0.33}$& $ 68.76^{+0.33}_{-0.39} $ & $ 68.74\pm 0.40$& $68.77\pm 0.39 $ \\
$\sigma_8$       & $0.7921\pm 0.0078 $ &$0.8058\pm 0.0069$ & $ 0.8078\pm 0.0075$ & $0.8075^{+0.0054}_{-0.0062} $& $ 0.8074^{+0.0056}_{-0.0064}$ \\
$ S_8$       & $ 0.8068^{+0.0079}_{-0.010}$ &$ 0.8080^{+0.0084}_{-0.0097}$ & $ 0.8087\pm 0.0094$ & $ 0.8086\pm 0.0094$& $ 0.8080\pm 0.0085$ \\
$ \Omega_m$       & $0.3112^{+0.0040}_{-0.0059} $ & $ 0.3017^{+0.0036}_{-0.0049}$& $0.3006^{+0.0045}_{-0.0038} $ & $0.3008\pm 0.0049 $& $ 0.3004\pm 0.0045$ \\
\hline
\hline

$ 10^{2}\,a_t$              & $ 3.473^{+0.076}_{-0.31}$ &$3.379^{+0.058}_{-0.21}$ & $3.383^{+0.060}_{-0.22} $ & $ 3.36^{+0.12}_{-0.20}$& $3.358^{+0.052}_{-0.19} $ \\

\hline 
    \hline
    Total $\chi^2 $ & 3853.8996 & 3846.3778&3846.0869 &3845.8107 & 3846.0881\\
     $\Delta \chi^2 $ & 7.8223 &0.3005 &0.0096 & -0.2666& 0.0108\\

\hline
\hline

\end{tabular}}
 
 \caption{ The mean $\pm 1 \sigma$ constraints on the cosmological parameters inferred from the \textit{Planck} 2018 CMB data (including \texttt{TTEETE} + lensing), BAO, SNe, DES-Y1 and weak-lensing data for $\Lambda$CDM and Canonical Emergent DE scenario with different values of $n$ varying from 1 to 6. }
 \label{tab:CBSDKcs1}
\end{table}

\section{Clustering Emergent Dark Energy } \label{apn:cs0}

    \subsection{ Planck TTTEEE only}

    \begin{table}[H]
    Constraints from Planck 2018 CMB for Clustering Emergent Dark Energy
   \vspace{1 em}  \\
\centering
\scalebox{1}{
\begin{tabular}{|c|c|c|c|c|c|}

\hline
\hline

    Parameter         &$n=1$ & $n=2$  & $n=3$ & $n=4$&$n=6$ \\
\hline
\hline
$ H_0 $     & $66.63\pm 0.65 $ & $68.62\pm 0.93 $& $ 68.90^{+0.82}_{-0.73}$&$ 69.02^{+0.84}_{-0.74}$ &$ 68.89^{+0.90}_{-0.72}$ \\
$\sigma_8$  &$ 0.8371^{+0.0080}_{-0.012}$  & $0.8633^{+0.0092}_{-0.012} $ &$0.8638^{+0.0089}_{-0.012} $ & $ 0.8649^{+0.0080}_{-0.013}$ &$ 0.8632^{+0.0086}_{-0.012}$ \\
$ S_8$ & $ 0.863^{+0.016}_{-0.019}$ &$ 0.864^{+0.018}_{-0.022}$ &$0.861^{+0.016}_{-0.020} $ &$ 0.859^{+0.017}_{-0.019}$ &$ 0.860^{+0.016}_{-0.019}$ \\
$ \Omega_m$ & $ 0.3189^{+0.0086}_{-0.0099}$ &$ 0.3009^{+0.0095}_{-0.012}$ &$0.2979^{+0.0087}_{-0.011} $ &$0.2964^{+0.0090}_{-0.011} $ &$ 0.2980^{+0.0087}_{-0.012}$ \\
\hline
\hline

$ 10^{2}\,a_t$        & $ 3.335^{+0.051}_{-0.17}$ & $3.274^{+0.029}_{-0.11} $&$ 3.289^{+0.052}_{-0.13}$ & $ 3.283^{+0.032}_{-0.12}$&$ 3.268^{+0.039}_{-0.10}$ \\

\hline 
    \hline
    Total $\chi^2 $ & 2764.9465 &2766.3194 & 2766.6314& 2766.6984& 2766.6043\\
     $\Delta \chi^2 $ & 0.7765 & 2.1494&2.4614 & 2.5284& 2.4343\\
\hline
\hline
\end{tabular}}
 
 \caption{ The mean $\pm 1 \sigma$ constraints on the cosmological parameters inferred from the \textit{Planck} 2018 CMB data (including \texttt{TTEETE}) for $\Lambda$CDM and Clustering Emergent DE scenario with different values of $n$ varying from 1 to 6. }
 \label{tab:Planckcs0}
\end{table}

    \subsection{ Planck + Lensing + BAO + $H_0$}
 \begin{table}[H]  
 Constraints from Planck 2018 + Lensing + BAO + SN +$H_0$ for Clustering Emergent Dark Energy
   \vspace{1 em}  \\
\centering
\scalebox{1}{
\begin{tabular}{|c|c|c|c|c|c|}

\hline
\hline

    Parameter         &$n=1$ & $n=2$  & $n=3$ & $n=4$&$n=6$ \\

\hline
\hline
$ H_0 $          & $67.34^{+0.69}_{-0.29} $ &$68.51^{+0.70}_{-0.38} $ & $68.76^{+0.63}_{-0.46} $ & $68.74^{+0.71}_{-0.43} $& $ 68.65^{+0.41}_{-0.35}$ \\
$\sigma_8$       & $0.8365^{+0.0063}_{-0.012} $ & $ 0.8551^{+0.0059}_{-0.0091}$& $0.8540^{+0.0048}_{-0.011} $ & $0.8145^{+0.0068}_{-0.010}$& $.8581^{+0.0061}_{-0.0076}$ \\
$ S_8$       & $0.8487^{+0.0093}_{-0.019}$ & $0.8578^{+0.0087}_{-0.016} $ & $0.8531^{+0.0083}_{-0.017} $ & $0.814^{+0.011}_{-0.017} $& $0.8591^{+0.0089}_{-0.010}$ \\
$ \Omega_m$       & $0.3088^{+0.0039}_{-0.0095} $ & $0.3020^{+0.0045}_{-0.0090} $ & $0.2995^{+0.0055}_{-0.0079} $ & $0.2999^{+0.0049}_{-0.0089} $& $ 0.3007^{+0.0046}_{-0.0057} $ \\
\hline
\hline

$ 10^{2}\,a_t$              & $ 3.375^{+0.049}_{-0.21}$ &$3.257^{+0.015}_{-0.094} $ & $3.275^{+0.014}_{-0.11} $ & $3.284^{+0.031}_{-0.12} $& $3.244^{+0.022}_{-0.080}$ \\

\hline 
    \hline
    Total $\chi^2 $ & 3860.9091 &3853.7403& 3852.5179&3855.1856 &3857.67 \\
     $\Delta \chi^2 $ & 6.8811 &-0.2877 &-1.5101 &1.1576 & 3.642\\

\hline
\hline

\end{tabular}}
 
 \caption{ The mean $\pm 1 \sigma$ constraints on the cosmological parameters inferred from the \textit{Planck} 2018 CMB data (including \texttt{TTEETE} + lensing), BAO, SNe and SH0ES for $\Lambda$CDM and Clustering Emergent DE scenario with different values of $n$ varying from 1 to 6. }
 \label{tab:CBSHcs0}
\end{table}
    \subsection{ Planck + Lensing + BAO + $H_0$ + DES-Y1}

\begin{table}[H]
\centering
Constraints from Planck 2018 + Lensing + BAO + SN +$H_0$ + DES
   \vspace{1 em}  \\

\scalebox{1}{
\begin{tabular}{|c|c|c|c|c|c|c|}

\hline
\hline

    Parameter         &$n=1$ & $n=2$  & $n=3$ & $n=4$&$n=6$ & $n$ open\\
\hline
\hline
$ H_0 $          & $67.61^{+0.60}_{-0.29}$ &$69.00\pm 0.51 $ & $68.88^{+0.48}_{-0.36}$ & $68.96^{+0.43}_{-0.27}$& $69.04^{+0.52}_{-0.42}$ & $68.93^{+0.50}_{-0.30}$\\
$\sigma_8$       & $0.8286^{+0.0061}_{-0.0088}$ &$0.8505^{+0.0057}_{-0.0083}$ & $0.8463^{+0.0058}_{-0.0072}$ & $0.8527\pm 0.0075$& $0.8534^{+0.0048}_{-0.0075}$ & $0.830^{+0.019}_{-0.014}$ \\
$ S_8$       & $0.8355^{+0.0082}_{-0.014}$ &$0.844\pm 0.023$ & $0.8429^{+0.0078}_{-0.011} $ & $0.8478^{+0.0072}_{-0.011} $& $0.8471^{+0.0087}_{-0.011}$ & $0.825\pm 0.027$\\
$ \Omega_m$       & $0.3051^{+0.0035}_{-0.0082}$ &$0.2956\pm 0.0060 $ & $0.2976^{+0.0043}_{-0.0061}$ & $ 0.2966^{+0.0030}_{-0.0056}$& $0.2957^{+0.0049}_{-0.0066} $ & $0.2967^{+0.0037}_{-0.0058}$\\
\hline
\hline
$ 10^{2}\,a_t$              & $3.301^{+0.035}_{-0.14} $ &$3.228^{+0.011}_{-0.065}$ & $3.220^{+0.011}_{-0.058} $ & $3.218^{+0.012}_{-0.056} $& $ 3.221^{+0.011}_{-0.060}$ & $3.244^{+0.020}_{-0.081}$\\
$n$ & 1 &2 &3&4&6& $3.35^{+0.78}_{-0.011}$ \\

\hline 
    \hline
    Total $\chi^2 $ & 3865.0585  &3860.6482 & 3859.5398& 3861.9733& 3863.1704 & 3855.541 \\
     $\Delta \chi^2 $ & 9.3301 & 4.9198& 3.8114& 6.2449& 7.442 & 0.1874 \\

\hline
\hline

\end{tabular}}
 
 \caption{ The mean $\pm 1 \sigma$ constraints on the cosmological parameters inferred from the \textit{Planck} 2018 CMB data (including \texttt{TTEETE} + lensing), BAO, SNe, SH0ES and DES-Y1 for $\Lambda$CDM and Clustering Emergent DE scenario with different values of $n$ varying from 1 to 6. }
 \label{tab:CBSHDcs0}
\end{table}
    \subsection{ Planck + Lensing + BAO + $H_0$ + DES-Y1 + HSC + KIDS}

\begin{table}[H]
Constraints from Planck 2018 + Lensing + BAO + SN +$H_0$ + DES + KIDS + HSC
   \vspace{1 em}  \\
\centering
\scalebox{1}{
\begin{tabular}{|c|c|c|c|c|c|}

\hline
\hline

    Parameter         &$n=1$ & $n=2$  & $n=3$ & $n=4$&$n=6$ \\

\hline
\hline
$ H_0 $          & $67.61^{+0.65}_{-0.43}$ &$68.99^{+0.56}_{-0.27}$ & $68.98^{+0.66}_{-0.38}$ & $69.07^{+0.43}_{-0.39}$& $69.41^{+0.38}_{-0.56} $ \\
$\sigma_8$       & $0.8242^{+0.0052}_{-0.0083}$ &$ 0.844\pm 0.019$ & $0.8418\pm 0.0088$ & $0.8092^{+0.0056}_{-0.0083} $& $0.8496^{+0.0054}_{-0.0071} $ \\
$ S_8$       & $0.8307^{+0.0086}_{-0.015}$ &$0.8376^{+0.0070}_{-0.011}$ & $0.8364^{+0.0087}_{-0.012}$ & $ 0.8030^{+0.0084}_{-0.011}$& $0.837\pm 0.033 $ \\
$ \Omega_m$       & $0.3049^{+0.0053}_{-0.0089}$ &$0.2956^{+0.0031}_{-0.0070}$ & $0.2962^{+0.0045}_{-0.0080}$ & $0.2954^{+0.0045}_{-0.0054} $& $0.2909^{+0.0067}_{-0.0046}$ \\
\hline
\hline
$ 10^{2}\,a_t$              & $3.260^{+0.018}_{-0.098}$ &$3.210^{+0.011}_{-0.049}$ & $3.2105^{+0.0096}_{-0.049}$ & $3.254^{+0.017}_{-0.091}$& $3.2151^{+0.0054}_{-0.053}$ \\

\hline 
    \hline
    Total $\chi^2 $ &  3871.3678 & 3869.4694 & 3867.4459&3859.2281&3871.5003 \\
     $\Delta \chi^2 $ & 11.556 &9.6576 & 7.6341& -0.5837&11.6885 \\

\hline
\hline

\end{tabular}}
 
 \caption{ The mean $\pm 1 \sigma$ constraints on the cosmological parameters inferred from the \textit{Planck} 2018 CMB data (including \texttt{TTEETE} + lensing), BAO, SNe, SH0ES, DES-Y1 and weak-lensing data for $\Lambda$CDM and Clustering Emergent DE scenario with different values of $n$ varying from 1 to 6. }
 \label{tab:CBSHDKcs0}
\end{table}

    \subsection{ Planck + Lensing + BAO + DES-Y1 + HSC + KIDS}

\begin{table}[H]
Constraints from Planck 2018 + Lensing + BAO + SN +$H_0$ + DES 
   \vspace{1 em}  \\
\centering
\scalebox{0.9}{
\begin{tabular}{|c|c|c|c|c|c|}

\hline
\hline

    Parameter         &$n=1$ & $n=2$  & $n=3$ & $n=4$&$n=6$ \\
\hline
\hline
$ H_0 $          & $67.51^{+0.41}_{-0.33}$ &$ 68.87^{+0.44}_{-0.31}$ & $ 68.97\pm 0.44$ & $68.82\pm 0.37 $& $68.90^{+0.46}_{-0.37}$ \\
$\sigma_8$       & $0.8234^{+0.0059}_{-0.0068}$ & $0.8455^{+0.0057}_{-0.010}$& $0.8422^{+0.0052}_{-0.0070}$ & $0.8081^{+0.0060}_{-0.0077}$& $0.8482\pm 0.0090 $ \\
$ S_8$       & $ 0.8317^{+0.0074}_{-0.010}$ & $ 0.8413^{+0.0058}_{-0.013}$& $ 0.8369^{+0.0084}_{-0.0096} $ & $ 0.8059^{+0.0080}_{-0.0098}$& $0.8443^{+0.0074}_{-0.011}$ \\
$ \Omega_m$       & $0.3061^{+0.0041}_{-0.0057}$ &$0.2970^{+0.0038}_{-0.0058} $ & $ 0.2963^{+0.0044}_{-0.0051}$ & $0.2984^{+0.0040}_{-0.0048}$& $0.2973^{+0.0043}_{-0.0059}$ \\
\hline
\hline
$ 10^{2}\,a_t$              & $ 3.253^{+0.021}_{-0.091} $ &$3.2117^{+0.0082}_{-0.049} $ & $3.2071^{+0.0099}_{-0.046}$ & $3.241^{+0.030}_{-0.078} $& $3.2060^{+0.0072}_{-0.044}$ \\

\hline 
    \hline
    Total $\chi^2 $ & 3854.213 & 3859.4984&  3857.8257 &3848.0467& 3861.4739\\
     $\Delta \chi^2 $ & 8.1357 & 13.4211& 11.7484& 1.9694& 15.3966\\

\hline
\hline

\end{tabular}}
 
 \caption{ The mean $\pm 1 \sigma$ constraints on the cosmological parameters inferred from the \textit{Planck} 2018 CMB data (including \texttt{TTEETE} + lensing), BAO, SNe, DES-Y1 and weak-lensing data for $\Lambda$CDM and Clustering Emergent DE scenario with different values of $n$ varying from 1 to 6. }
 \label{tab:CBSDKcs0}
\end{table}

\section{Non canonical Emergent Dark Energy} \label{apn:csopen}

   \subsection{ Planck TTTEEE only}
\begin{table}[H]
Constraints from Planck 2018 
   \vspace{1 em}  \\
\centering
\scalebox{0.9}{
\begin{tabular}{|c|c|c|c|c|c|}

\hline
\hline

    Parameter         &$n=1$ & $n=2$  & $n=3$ & $n=4$&$n=6$ \\
\hline
\hline
$ H_0 $          & $66.18\pm 0.62$ &$ 68.30\pm 0.70$ & $68.47^{+0.66}_{-0.74} $ & $68.48\pm 0.74$& $68.47\pm 0.77$ \\
$\sigma_8$       & $0.7997^{+0.0077}_{-0.0090}$ & $ 0.8132^{+0.0078}_{-0.0091}$& $0.8147^{+0.0075}_{-0.0098}$ & $0.8146^{+0.0077}_{-0.0097} $& $0.8143^{+0.0079}_{-0.0092}$ \\
$ S_8$       & $0.835\pm 0.017$ & $0.822\pm 0.017 $& $0.821\pm 0.017$ & $ 0.821\pm 0.017$& $0.821\pm 0.017$ \\
$ \Omega_m$       & $0.3268\pm 0.0091$ & $0.3067\pm 0.0089 $& $0.3050\pm 0.0091 $ & $0.3050\pm 0.0092 $& $0.3051\pm 0.0094$ \\
\hline
\hline
$ 10^{2}\,a_t$              & $ 3.414^{+0.066}_{-0.25}$ & $3.375^{+0.056}_{-0.21}$& $3.372^{+0.053}_{-0.21} $ & $ 3.372^{+0.058}_{-0.21}$& $3.372^{+0.054}_{-0.21}$ \\

$ c_s^{2}$        & $0.64^{+0.29}_{-0.16}$ & $0.66^{+0.27}_{-0.16} $& $ 0.65 \pm 0.21 $ & $0.65^{+0.28}_{-0.16} $&$0.66^{+0.28}_{-0.15}$ \\

\hline 
    \hline
    Total $\chi^2 $ & 2765.6229 &2766.0841& 2765.7061& 2765.8519&2765.4524 \\
     $\Delta \chi^2 $ & 1.4529 &1.9141 & 1.5361& 1.6819& 1.2824\\

\hline
\hline

\end{tabular}}
 
 \caption{ The mean $\pm 1 \sigma$ constraints on the cosmological parameters inferred from the \textit{Planck} 2018 CMB data (including \texttt{TTEETE} ) data for $\Lambda$CDM and Non-canonical Emergent DE scenario with different values of $n$ varying from 1 to 6. }
 \label{tab:Planckopencs}
\end{table}

\subsection{Planck 2018 + Lensing + BAO + SN +$H_0$ }
\begin{table}[H]
Constraints from Planck 2018 + Lensing + BAO + SN +$H_0$ 
   \vspace{1 em}  \\
\centering
\scalebox{0.9}{
\begin{tabular}{|c|c|c|c|c|c|}

\hline
\hline

    Parameter         &$n=1$ & $n=2$  & $n=3$ & $n=4$&$n=6$ \\
\hline
\hline
$ H_0 $          & $67.52\pm 0.36$ & $68.72^{+0.55}_{-0.33}$ & $68.83^{+0.44}_{-0.50}$ & $68.79\pm 0.39$& $ 68.89\pm 0.53'$ \\
$\sigma_8$       & $0.8014^{+0.0069}_{-0.0088}$ & $0.8154^{+0.0067}_{-0.0086}$ & $0.8150^{+0.0065}_{-0.0078}$ & $0.8247^{+0.0064}_{-0.0076}$& $0.8155^{+0.0064}_{-0.0080}$ \\
$ S_8$       & $0.8121^{+0.0096}_{-0.011}$ & $0.8174^{+0.0086}_{-0.014}$ & $0.816\pm 0.011$ & $0.826\pm 0.013$& $0.8156^{+0.0089}_{-0.011}$ \\
$ \Omega_m$       & $0.3081\pm 0.0048$ & $0.3015^{+0.0038}_{-0.0067} $ & $ 0.3007\pm 0.0058$ & $ 0.3010\pm 0.0045$ & $0.3001^{+0.0044}_{-0.0055}$ \\
\hline
\hline
$ 10^{2}\,a_t$              & $3.502^{+0.097}_{-0.34}$ & $3.408^{+0.080}_{-0.24}$ & $3.417^{+0.077}_{-0.25}$ & $3.405^{+0.083}_{-0.23}$& $ 3.425^{+0.081}_{-0.26}$ \\

$ c_s^{2}$        & $< 0.665$ & $0.61^{+0.28}_{-0.20}$& $0.66^{+0.23}_{-0.19}$ & $0.61^{+0.19}_{-0.25}$&$0.65^{+0.27}_{-0.17}$ \\

\hline 
    \hline
    Total $\chi^2 $ & 3860.7531 &3852.0743 &3851.7264 & 3851.8387&3851.6283\\
     $\Delta \chi^2 $ & 6.7251 &-1.9537 & -2.3016& -2.1893& -2.3997\\

\hline
\hline

\end{tabular}}
 
 \caption{ The mean $\pm 1 \sigma$ constraints on the cosmological parameters inferred from the \textit{Planck} 2018 CMB data (including \texttt{TTEETE} + lensing), BAO, SNe, SH0ES data for $\Lambda$CDM and non-canonical Emergent DE scenario with different values of $n$ varying from 1 to 6.  }
 \label{tab:CBSHopencs}
\end{table}
    
    \subsection{ Planck + Lensing + BAO + $H_0$ + DES-Y1}

\begin{table}[H]
\centering
Constraints from Planck 2018 + Lensing + BAO + SN +$H_0$ + DES
   \vspace{1 em}  \\

\scalebox{0.9}{
\begin{tabular}{|c|c|c|c|c|c|c|}

\hline
\hline

    Parameter         &$n=1$ & $n=2$  & $n=3$ & $n=4$&$n=6$ & $n$ open\\
\hline
\hline
$ H_0 $          & $67.65^{+0.29}_{-0.43}$ &$68.94^{+0.37}_{-0.48}$ & $68.93^{+0.44}_{-0.36}$ & $69.14^{+0.41}_{-0.56}$& $69.02^{+0.38}_{-0.46}$ & $68.96\pm 0.40$\\
$\sigma_8$       & $0.7994^{+0.0061}_{-0.0096}$ &$0.8125^{+0.0063}_{-0.0082}$ & $0.8129^{+0.0063}_{-0.0085}$ & $0.8126^{+0.0061}_{-0.0077}$& $0.8134^{+0.0065}_{-0.0079}$ & $0.8125^{+0.0059}_{-0.0079}$\\
$ S_8$       & $0.8078^{+0.0098}_{-0.011}$ &$0.811\pm 0.011$ & $0.8119^{+0.0098}_{-0.011}$ & $0.8086^{+0.0089}_{-0.010}$& $0.8112^{+0.0087}_{-0.011}$ & $ 0.8108^{+0.0084}_{-0.010}$\\
$ \Omega_m$       & $0.3064^{+0.0051}_{-0.0041}$ & $0.2988^{+0.0052}_{-0.0046}$& $0.2992^{+0.0042}_{-0.0052}$ & $0.2971^{+0.0059}_{-0.0052}$& $0.2984\pm 0.0056$ & $0.2988\pm 0.0046$\\
\hline
\hline
$ 10^{2}\,a_t$              & $3.510^{+0.088}_{-0.35}$ &$ 3.415^{+0.064}_{-0.25}$ & $3.392^{+0.074}_{-0.23}$ & $3.447^{+0.092}_{-0.27}$& $3.407^{+0.070}_{-0.24}$ & $3.400^{+0.069}_{-0.23}$\\

$ c_s^{2}$        & $0.58\pm 0.23$ & $0.65^{+0.30}_{-0.14}$& $0.65^{+0.28}_{-0.16}$ & $0.69^{+0.20}_{-0.18}$&$0.64^{+0.32}_{-0.14}$ & $0.66^{+0.27}_{-0.15}$\\
$n$ & 1&2&3&4&6&$ > 3.07$ \\

\hline 
    \hline
    Total $\chi^2 $ & 3868.9616 &3853.8864  &  3853.4742& 3853.0872& 3853.4632 & 3852.9647\\
     $\Delta \chi^2 $ &13.2332  & -1.842&-2.2542 &-2.6412 & -2.2652 & -2.7637\\

\hline
\hline

\end{tabular}}
 
 \caption{ The mean $\pm 1 \sigma$ constraints on the cosmological parameters inferred from the \textit{Planck} 2018 CMB data (including \texttt{TTEETE} + lensing), BAO, SNe, SH0ES and DES-Y1 data for $\Lambda$CDM and non-canonical Emergent DE scenario with different values of $n$ varying from 1 to 6.  }
 \label{tab:CBSHDopencs}
\end{table}
    
    \subsection{ Planck + Lensing + BAO + $H_0$ + DES-Y1 + HSC + KIDS}

\begin{table}[H]
Constraints from Planck 2018 + Lensing + BAO + SN +$H_0$ + DES + KIDS + HSC
   \vspace{1 em}  \\
\centering
\scalebox{0.8}{
\begin{tabular}{|c|c|c|c|c|c|}

\hline
\hline

    Parameter         &$n=1$ & $n=2$  & $n=3$ & $n=4$&$n=6$ \\
\hline
\hline
$ H_0 $          & $67.66^{+0.48}_{-0.36}$ & $69.03\pm 0.70$& $69.20^{+0.42}_{-0.54}$ & $69.12^{+0.40}_{-0.47}  $& $69.24^{+0.27}_{-0.48}$ \\
$\sigma_8$       & $0.7957^{+0.0058}_{-0.0081}$ & $0.8099^{+0.0049}_{-0.0085}$& $0.8126^{+0.0045}_{-0.011}$ & $0.8102^{+0.0053}_{-0.0082}$& $0.8103^{+0.0055}_{-0.0083}$ \\
$ S_8$       & $0.8037^{+0.0076}_{-0.012}$ &$0.8068^{+0.0081}_{-0.010}$ & $0.8074^{+0.0072}_{-0.013}$ & $0.8060^{+0.0080}_{-0.011} $& $0.804\pm 0.010$ \\
$ \Omega_m$       & $0.3061^{+0.0043}_{-0.0062}$ &$0.2977\pm 0.0074$ & $0.2962\pm 0.0060$ & $0.2969\pm 0.0063$& $0.2957^{+0.0053}_{-0.0035} $ \\
\hline
\hline
$ 10^{2}\,a_t$              & $3.486^{+0.071}_{-0.32}$ &$3.422^{+0.066}_{-0.25}$ & $3.454^{+0.072}_{-0.29}$ & $3.407^{+0.052}_{-0.24}$& $ 3.424^{+0.053}_{-0.26}$ \\

$ c_s^{2}$        & $0.59^{+0.20}_{-0.23}$ & $0.69^{+0.23}_{-0.15}$& $> 0.592 $ & $0.681^{+0.31}_{-0.089} $&$0.69^{+0.29}_{-0.10}$ \\

\hline 
    \hline
    Total $\chi^2 $ & 3872.2972 & 3857.2175&3857.2294&3856.8636&3856.7384\\
     $\Delta \chi^2 $ & 12.4852 & -2.5943& -2.5824& -2.9482& -3.0734\\

\hline
\hline

\end{tabular}}
 
 \caption{ The mean $\pm 1 \sigma$ constraints on the cosmological parameters inferred from the \textit{Planck} 2018 CMB data (including \texttt{TTEETE} + lensing), BAO, SNe, SH0ES, DES-Y1 and weak-lensing data for $\Lambda$CDM and non-canonical Emergent DE scenario with different values of $n$ varying from 1 to 6.  }
 \label{tab:CBSHDKopencs}
\end{table}
    
    \subsection{ Planck + Lensing + BAO + DES-Y1 + HSC + KIDS}

\begin{table}[H]
Constraints from Planck 2018 + Lensing + BAO + SN + DES + KIDS + HSC 
   \vspace{1 em}  \\
\centering
\scalebox{0.9}{
\begin{tabular}{|c|c|c|c|c|c|}

\hline
\hline

    Parameter         &$n=1$ & $n=2$  & $n=3$ & $n=4$&$n=6$ \\
\hline
\hline
$ H_0 $          & $67.35\pm 0.33$ &$68.69\pm 0.39$ & $ 68.77\pm 0.38$ & $68.81^{+0.34}_{-0.40}$& $68.79\pm 0.40 $ \\
$\sigma_8$       & $0.7947^{+0.0055}_{-0.0071}$ &$0.8083^{+0.0054}_{-0.0066} $ & $0.8091^{+0.0058}_{-0.0070}$ & $ $& $0.8091^{+0.0054}_{-0.0068}$ \\
$ S_8$       & $ 0.8076^{+0.0082}_{-0.0093}$ &$ 0.8100^{+0.0080}_{-0.0090}$ & $0.8100^{+0.0079}_{-0.0094} $ & $0.8111^{+0.0082}_{-0.011}$& $0.8098^{+0.0082}_{-0.0093}$ \\
$ \Omega_m$       & $0.3099\pm 0.0044$ &$0.3013\pm 0.0046$ & $0.3007\pm 0.0045$ & $0.3004\pm 0.0047$& $0.3006\pm 0.0047$ \\
\hline
\hline
$ 10^{2}\,a_t$              & $3.419^{+0.068}_{-0.25}$ & $3.357^{+0.057}_{-0.19}$& $ 3.342^{+0.065}_{-0.18}$ & $3.350^{+0.047}_{-0.18}$& $3.333^{+0.043}_{-0.17}$ \\

$ c_s^{2}$        & $ 0.62^{+0.20}_{-0.23}$ & $ > 0.585$& $ 0.68^{+0.25}_{-0.14}$ & $ > 0.582$&$ > 0.595$ \\

\hline 
    \hline
    Total $\chi^2 $ & 3853.5801  & 3845.4901&3845.5449& 3845.2508&3845.3574\\
     $\Delta \chi^2 $ & 7.5028 & -0.5872& -0.5324& -0.8265& -0.7199\\

\hline
\hline

\end{tabular}}
 
 \caption{ The mean $\pm 1 \sigma$ constraints on the cosmological parameters inferred from the \textit{Planck} 2018 CMB data (including \texttt{TTEETE} + lensing), BAO, SNe, DES-Y1 and weak-lensing data for $\Lambda$CDM and non-canonical Emergent DE scenario with different  values of $n$ varying from 1 to 6.  }
 \label{tab:CBSDKopencs}
\end{table}

\end{document}